\documentclass[twocolumn]{jpsj3}
\usepackage{graphicx,amsmath,amssymb,bm}
\usepackage{braket}

\allowdisplaybreaks 

\usepackage{color}
\definecolor{Green}{rgb}{0,0.7,0}

\newcommand{\ET}{ $\alpha$-(BEDT-TTF)$_2$I$_3$}

\newcommand{\bk}{ \bm{k}}
\newcommand{\bkD}{ \bm{k}_{\rm D}}
\newcommand{\eD}{ \epsilon_{\rm D}}
\newcommand{\ep}{ \epsilon }
\newcommand{\hn}{ \hat{n}}
\newcommand{\tV}{ \tilde{V}}

\newcommand{\g}{ \gamma }
\newcommand{\bq}{ \bm{q}}

\begin{document}

\title{
Anomalous Conductivity of Two-dimensional Dirac Electrons 
in Organic Conductor under Pressure}
\author{
Yoshikazu Suzumura\thanks{E-mail: suzumura@s.phys.nagoya-u.ac.jp}
$^{1}$
 and Masao Ogata
$^{2}$
}
\inst{
$^1$Department of Physics, Nagoya University,  Nagoya 464-8602, Japan \\
$^2$Department of Physics, University of Tokyo, Bunkyo, Tokyo 113-0033, Japan
 \\
}


\abst{
The electric conductivity of 
Dirac electrons in the organic conductor  \ET \; 
 [BEDT-TTF  = bis(ethylenedithio)tetrathiafulvalene]
under pressure 
  has been examined using  a two-dimensional tight-binding (TB) model
 with both  impurity  and electron--phonon (e--p) scatterings.  
 We study an anomalous temperature dependence of the conductivity,  
 which  shows   a crossover 
 from   $\sigma_{x} < \sigma_{y}$ at low temperatures [region (I)] 
 to  $\sigma_{x} > \sigma_{y}$  at high temperatures [region (II)]. 
$\sigma_y$ and $\sigma_x$ are diagonal conductivities 
 parallel and perpendicular to  a stacking axis of  molecules, respectively.
The effect of   Dirac cone tilting is dominant in  region (I), whereas 
 the anisotropy of 
  the velocity of the Dirac cone  is dominant in  region (II). 
Such  behavior is  further examined  by
  calculating the  deviation of principal axes 
    due to the  off-diagonal conductivity $\sigma_{xy}$ 
     and   a nearly constant  
      conductivity at high temperatures  
        due to the e--p scattering, which is   
   the extension of  the previous result of  the simple  two-band model 
   [Phys. Rev. B {\bf 98},161205 (2018)].
The relevance to  experiments of   organic conductors is discussed. 
  }


\maketitle

\section{Introduction} 

 Since the discovery of two-dimensional massless Dirac fermions,
\cite{Novoselov2005_Nature438}
  extensive studies  
 have been performed  on various materials. 
Among them, noticeable phenomena of Dirac electrons
 in  molecular crystals\cite{Kajita_JPSJ2014} have been studied  
in the organic conductor
\ET \cite{Mori1984} [BEDT-TTF=bis(ethylenedithio)tetrathiafulvalene].
 After  noting that the density of states (DOS) 
vanishes linearly at the Fermi energy,
\cite{Kobayashi2004} 
the two-dimensional  Dirac cone with a zero-gap state (ZGS) 
\cite{Katayama2006_JPSJ75} 
was found  using  a tight-binding (TB) model, where  
   transfer energies are estimated from 
 the extended H\"uckel method.\cite{Kondo2005} 
 The existence of such Dirac cone  was verified by  first-principles DFT calculation,
\cite{Kino2006} which has been  
 used for studying further \ET\; under hydrostatic 
  pressure.\cite{Katayama_EPJ}

There are   common features among organic conductors with  
 isostructure salts,~\cite{Inokuchi1993,Inokuchi1995_BCSJ68}
$\alpha$-D$_2$I$_3$ (D = ET, STF, and  BETS), where 
ET = BEDT-TTF, 
 STF = bis(ethylenedithio)diselenadithiafuluvalene), 
 and 
BETS =  bis(ethylenedithio)tetraselenafulvalene.
These salts display  an energy band  with a Dirac cone,
\cite{Katayama2006_JPSJ75,Kondo2009,Morinari2014,Naito2020} 
 and  the resistivity at high temperatures
 shows a nearly constant behavior,
\cite{
Inokuchi1993,Inokuchi1995_BCSJ68,Kajita1992,Tajima2000,Tajima2002,Tajima2007,
Liu2016
}
 whereas  the  conventional metal shows   
  the  linearly increasing one.
Such unconventional behavior  was also observed 
in Dirac electrons with nodal line semimetals of  single-component molecular conductors 
\cite{
Kato_JACS,Kato2017_JPSJ,Suzumura2018,Zhou2019,Kato2020_JPSJ
}
Thus, it has been considered  that 
 the nearly constant behavior of  resistivity at high temperatures 
 is attributable  
 to the intrinsic property of  Dirac electrons.

Thus, the conductivity of Dirac electrons has been studied in simple models.
 A two-band model  shows that the static conductivity 
at absolute zero temperature  remains finite with a universal value, i.e., independent of the magnitude of impurity scattering  
 owing to a quantum effect.\cite{Ando1998} 
The conductivity increases with increasing  
   doping concentration.
The effect of   Dirac cone tilting at absolute zero 
 was studied previously.\cite{Suzumura_JPSJ_2014}  
 It   provides  the anisotropic conductivity   
 and  the deviation of the current from the applied electric field.
\cite{Suzumura_JPSJ_2014}
At finite temperatures, on the other hand, 
the conductivity  depends on the magnitude 
 of the impurity scattering, $\Gamma$, which is proportional to the inverse of 
 the life-time by the disorder. 
With increasing temperature ($T$), the conductivity remains unchanged 
for $T \ll \Gamma$,
 whereas it increases for $\Gamma \ll T$.\cite{Neto2006} 
Noting that  $\Gamma \sim$ 0.0003 eV for organic conductors,\cite{Kajita_JPSJ2014}
 a  monotonic increase in 
the conductivity at finite temperature $T > 0.0005$ eV
 is expected.
However  the measured conductivity (or resistivity) on the above organic conductor 
 shows  an almost constant behavior at high temperatures. 
 To comprehend such an exotic  phenomenon, 
  the  acoustic phonon scatterings has been  proposed 
  as  a possible mechanism, which was studied 
  using  a simple two-band model of the Dirac cone  without tilting.
~\cite{Suzumura_PRB_2018}
Although  the scattering  shows 
     a  reasonable  suppression  of the conductivity  at high temperatures, 
  it is unclear if  such a model  sufficiently explains   
 the conductivity of the actual organic conductor, where the Dirac cone shows 
 deviation  from the linear spectrum  
 and then the calculation must be performed beyond  the model. 

In this paper, we study the anomalous temperature dependence 
of the anisotropic conductivity of 
$\alpha$-(BEDT-TTF)$_2$I$_3$ using the TB model~\cite{Kobayashi2007,Goerbig2008,Kobayashi2008,Katayama_EPJ}
 and acoustic phonons. Thus, the present paper is a natural extension of the previous study~\cite{Suzumura_PRB_2018}
 in which a simplified model was used without the details of the lattice structure 
of $\alpha$-(BEDT-TTF)$_2$I$_3$. By using the TB model,
we can  automatically take into account the effects of tilting and anisotropy discussed at absolute zero~ \cite{Suzumura_JPSJ_2014}
  to explore the temperature dependence.
 It will be shown that the presence of  acoustic phonons gives rise to conductivity being nearly constant at high temperatures.
We will also demonstrate 
 a crossover of the conductivity from a quantum regime at low temperatures  
to a classical regime at high temperatures by focusing on a competition 
 between  the tilted Dirac cone and the anisotropic velocity of the cone. 
We use the TB model with transfer energies of \ET\; 
under pressure, which are well known compared with those of 
other  isostructural salts, BETS and STF. 
Since the tilted  Dirac cone is also obtained for BETS~\cite{Morinari2014}  
and STF~\cite{Naito2020}, 
 it is expected that the present result 
    provides a common feature for these salts, 
i.e.,   the  nearly constant conductivity  at high temperatures.

The paper is organized as follows.
 In Sect. 2, the model and formulation are given  
 for both the uniaxial  and hydrostatic pressures, 
where the latter case is examined by adding  site potential. 
In Sect. 3, after examining the chemical potential and density of states (DOS),
 the conductivity is calculated 
 for \ET\;  under both  uniaxial and 
  hydrostatic pressures. 
 Section  4 is devoted to summary and discussion.

\section{Model and Formulation}

\begin{figure}
  \centering
\includegraphics[width=7cm]{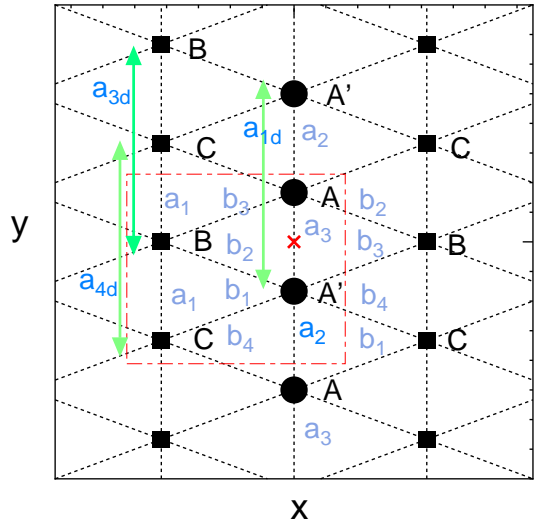}
     \caption{(Color online)
Crystal structure, where 
there are four molecules A, A', B, and C in the unit cell (dot-dashed line), 
which forms a square lattice.
Transfer energies are shown  by $a_1, \cdots, b_4$ 
 for the nearest neighbor (NN) sites and
$a_{1d}$, $a_{3d}$, and $a_{4d}$ for the next-nearest neighbor (NNN) sites. 
The cross denotes an inversion center between A and A'. 
}
\label{fig1}
\end{figure}

We consider a two-dimensional Dirac electron system per spin,
 which is given by 
\begin{equation}
H_{\rm total}= H_0 + H_{1} + H_{\rm p} +  H_{\rm e-p} +H_{imp} \; . 
\label{eq:H}
\end{equation}
 $H_0$ describes a TB model of 
 the  organic conductor \ET\;  consisting of four molecules per unit cell 
(Fig.~\ref{fig1}), where 
 $H_{1}$ describes a site potential obtained from a  mean field 
 of  short-range repulsive interactions.\cite{Katayama_EPJ}
 $H_{\rm p}$ and 
 $H_{\rm e-p}$ denote  an acoustic phonon and 
an  electron-phonon (e--p) interaction, respectively. 
$H_{\rm imp}$ is the impurity potential. 
The terms $H_0 + H_{\rm p} + H_{\rm e-p}$  provide 
the   Fr\"ohlich Hamiltonian 
\cite{Frohlich}     applied to the  present Dirac electron system.
The unit of the energy is taken as eV.
\subsection{Energy band}
First, we  derive the energy band for $H=H_0+H_1$ 
and the associated quantities.
A TB model, $H_0$, is expressed as 
\begin{eqnarray}
H_0 &=& \sum_{i,j = 1}^N \sum_{\alpha, \beta = 1}^4
 t_{i,j; \alpha,\beta} a^{\dagger}_{i,\alpha} a_{j, \beta} 
                          \nonumber \\
&=& \sum_{\bk}  \sum_{\alpha, \beta = 1}^4
 t_{\alpha, \beta}(\bk)  a^{\dagger}_{\alpha}(\bk) a_{\beta}(\bk) 
\; , 
\label{eq:Hij}
\end{eqnarray}
where   $a^{\dagger}_{i, \alpha}$ denotes a creation operator 
 of an electron 
 of molecule $\alpha$ 
 [= A(1), A'(2), B(3), and C(4)]  in the unit cell 
  at  the $i$-th lattice site.  
 $N$ is the total number of square lattice sites and 
 $t_{i,j; \alpha,\beta}$ are the transfer energies 
for  the nearest  neighbor (NN) and next-nearest 
 neighbor (NNN) sites, which are shown in Fig.~\ref{fig1}. 
A Fourier transform for the operator $a_{j,\alpha}$ 
 is given by   
 $a_{j,\alpha} = 1/N^{1/2} \sum_{\bk} a_{\alpha}(\bk) \exp[ i \bk \cdot \bm{r}_j]$, 
where $\bk = (k_x,k_y)$ and the lattice constant is taken as unity.
The quantity $H_1$ corresponds to   a site potential, 
 $V_{\alpha}$,  acting on 
 the $\alpha$ site, where 
  $V_{\rm A} = V_{\rm A'}$ due to an inversion symmetry 
  around the cross in Fig. \ref{fig1}.  
 The Hamiltonian $H_1$ is obtained as (Appendix A)
\begin{eqnarray}
H_1 & = &\sum_{\alpha} (V_{\alpha} - V_{\rm A})\hn_{\alpha} 
                   \nonumber \\
    & = & \tV_{\rm B} \hn_{\rm B} + \tV_{\rm C} \hn_{\rm C} \; , 
\label{eq:H1}
\end{eqnarray}
 where $\tV_{\alpha}$ denotes a  potential measured from that of the A site  
 and  $\hn_{\alpha} =  a^{\dagger}_{\alpha}(\bk) a_{\beta}(\bk)$.  
From Eqs.~(\ref{eq:Hij}) and (\ref{eq:H1}),
 $H$ is written as\cite{Katayama_EPJ} 
\begin{eqnarray}
  H = \sum_{\bk} \sum_{\alpha, \beta}  
      a^{\dagger}_{\alpha}(\bk) h_{\alpha,\beta} a_{\beta}(\bk)\; , 
\label{eq:H_total}
\end{eqnarray}
 with $h_{\alpha,\beta}$ being the matrix element (Appendix A)
and  is diagonalized as 
\begin{subequations}
\begin{eqnarray}
\label{eq:eq6a}
 H = \sum_{\bk} \sum_{\g}
  c_\g^{\dagger}(\bk) E_\g(\bk) c_\g(\bk) \; ,
\end{eqnarray}
where $E_1(\bk) > E_2(\bk) > E_3(\bk) > E_4(\bk)$  and 
\begin{eqnarray}
\label{eq:eigen_eq}
\sum_{\beta} h_{\alpha,\beta}(\bk) d_{\beta \g(\bk)}
   &=& E_{\g}(\bk) d_{\alpha \g} (\bk)  \; , \\
\label{eq:eq6b}
c_\g(\bk)& = &\sum_{\alpha} d_{\alpha\g}(\bk)  a_{\alpha}(\bk) \; .
\end{eqnarray}
\end{subequations}
The Dirac point ($\bkD$) is calculated  from  
\begin{eqnarray}
\label{eq:ZGS}
E_1(\bkD) = E_2(\bkD)= \eD \; .
\end{eqnarray}
 The ZGS is obtained when 
 $\eD$ becomes equal to  the chemical potential at $T = 0$. 

From $E_\g$, the local density  $n_{\alpha}$ 
including  both spin $\uparrow$ and $\downarrow$ 
   is calculated as   
\begin{eqnarray}
 n_{\alpha} & = &\frac{2}{N} \sum_{\bk} 
     \left< \hn_{\alpha}(\bk) \right>_{H} \nonumber \\ 
 & = &
 \frac{2}{N} \sum_{\bk}\sum_{\g} 
  d^*_{\alpha\g}(\bk) d_{\alpha \g}(\bk)f(E_\g(\bk)-\mu)
 \; ,
\label{eq:local_charge}
 \end{eqnarray}  
 which is determined self-consistently. 
 $n_{\rm A} = n_{\rm A'}$  owing  to transfer energies  
 being symmetric     with respect to  the inversion center   between   A and A'
 in Fig.~\ref{fig1}.
In Eq.~(\ref{eq:local_charge}), $f(\ep)= 1/(\exp[\ep/T]+1)$ with $T$ being temperature in the unit of eV 
 and $k_{\rm B }=1$.
The chemical potential $\mu$ is determined 
 from the three-quarter-filled condition, which is given by 
\begin{eqnarray}
  \frac{1}{N} \sum_{\bk} \sum_{\gamma}  f(E_{\gamma}(\bk)-\mu)=
 \int_{-\infty}^{\infty} {\rm d} \omega D(\omega) f(\omega) =  3 \; ,  
  \label{eq:mu}
\end{eqnarray}
where  
\begin{eqnarray}
D(\omega) &=& \frac{1}{N} \sum_{\bk} \sum_{\gamma}
 \delta (\omega - E_{\gamma}(\bk)) \; .
  \label{eq:dos}
\end{eqnarray}
$D(\omega)$ denotes DOS per spin and per unit cell, 
 which satisfies  $\int {\rm d} \omega D(\omega) = 4$.
Note that $ n_{\rm A} + n_{\rm A'} + n_{\rm B} + n_{\rm C} =6$
from Eq.~(\ref{eq:mu}).
We use $\mu (T)$ at finite $T$ and $\mu$ = $\mu(0)$ at $T$=0. 

In Eq.~(\ref{eq:H}), the third term denotes  
  the harmonic phonon   given by  
 $H_{\rm p}= \sum_{\bq} \omega_{\bq} b_{\bq}^{\dagger} b_{\bq}$ 
 with $\omega_{\bq} = v_s |\bq|$ and  $\hbar$ =1,  whereas
 the fourth term is  the e--p interaction  expressed 
as~\cite{Frohlich} 
\begin{equation}
 H_{\rm e-p} = \sum_{\bk, \g} \sum_{\bq}
   \alpha_{\bq} c_\g(\bk + \bq)^\dagger c_{\g}(\bk) \phi_{\bq} \; ,
\label{eq:H_e--p}
\end{equation}
 with 
 $\phi_{\bq} = b_{\bq} + b_{-\bq}^{\dagger}$.
  We introduce  a coupling constant $\lambda = |\alpha_{\bq}|^2/\omega_{\bq}$, 
  which becomes  independent of $|\bq|$  for small $|\bq|$. 
The e--p scattering is considered 
 within  the same band (i.e., intraband) 
 owing  to the energy conservation with $v \gg v_s$, where 
  $v \simeq 0.05$~\cite{Katayama_EPJ} 
 denotes the averaged velocity of the Dirac cone. 
The last term of Eq.~(\ref{eq:H}), $H_{\rm imp}$, denotes a normal  impurity 
 scattering, which is  
 introduced to obtain the finite conductivity 
 and to avoid the infinite conductivity 
in the presence of only the e--p interaction 
\cite{Holstein1964}. 

\subsection{Conductivity}
By using the component of the wave function $d_{\alpha \gamma}$ 
 in Eq.~(\ref{eq:eq6b}), we calculate 
 the conductivity   per spin and per site 
   as\cite{Katayama2006_cond}  
\begin{eqnarray}
\sigma_{\nu \nu'}(T) &=&  
  \frac{e^2 }{\pi \hbar N} 
  \sum_{\bk} \sum_{\gamma, \gamma'} 
  v^\nu_{\gamma \gamma'}(\bk)^* 
  v^{\nu'}_{\gamma' \gamma}(\bk) \nonumber \\
& &  \int_{- \infty}^{\infty} d \ep 
   \left( - \frac{\partial f(\ep) }{\partial \ep} \right)
    \nonumber \\
  \times & &\frac{\Gamma_\g}{(\ep - \xi_{\bk \gamma})^2 + \Gamma_\g^2} \times 
 \frac{\Gamma_{\g'}}{(\ep - \xi_{\bk \gamma'})^2 +  \Gamma_{\g'}^2}
  \; ,  \nonumber \\
  \label{eq:sigma}
\\
  v^{\nu}_{\gamma \gamma'}(\bk)& = & \sum_{\alpha \beta}
 d_{\alpha \gamma}(\bk)^* 
   \frac{\partial h_{\alpha \beta}}{\partial k_{\nu}}
 d_{\beta \gamma'}(\bk) \; ,
  \label{eq:v}
\end{eqnarray}
 where $\nu = x$ and $y$.
 $h = 2 \pi \hbar$ and $e$ denote  Planck's constant and electric charge, 
 respectively.  
 The quantity $\Gamma_\g$ denotes 
 the damping of the electron of the $\g$ band given by 
\begin{eqnarray}
\Gamma_{\g}  = \Gamma + \Gamma_{\rm ph}^{\g} \; ,
\end{eqnarray}
where the first term comes from the impurity scattering 
and the second term corresponding to  the phonon scattering 
 is given by~\cite{Suzumura_PRB_2018} (Appendix B)
\begin{subequations}
\begin{eqnarray}
  \Gamma_{\rm ph}^\g &=& C_0R \times T|\xi_{\g,\bk}|
  \; ,
 \label{eq:eq16a}
        \\ 
R &=& \frac{\lambda}{ \lambda_0}
 \; ,  
 \label{eq:eq16b} 
\end{eqnarray}
 \end{subequations}
with  $C_0 = 6.25\lambda_0/(2\pi v^2)$.  
 For $v \simeq 0.05$  and $\lambda_0/2\pi v = 0.1 $,
 we obtain $C_0 \simeq$ 12.5 (eV)$^{-1}$.
$R$  denotes a normalized e--p coupling constant. 
 
In the following, we denote $\sigma_x$, $\sigma_y$, and $\sigma_{xy}$ 
instead of $\sigma_{xx}(T)$, $\sigma_{yy}(T)$, and $\sigma_{xy}(T)$ 
 for simplicity. 
In terms of  
  $\sigma_x$,  
 $\sigma_y$, and  
 $\sigma_{x y}$,  
 the current $(j_x,j_y)$ obtained from a response to 
 an external electric field 
 $(E_x,E_y)$  is written as 
\begin{eqnarray}
\begin{pmatrix}
  j_x  \\
  j_y
\end{pmatrix}  
=
\begin{pmatrix}
  \sigma_x  & \sigma_{xy}  \\
  \sigma_{xy}  & \sigma_y 
\end{pmatrix}  
\begin{pmatrix}
  E_x  \\
  E_y
\end{pmatrix} \; . 
 \label{eq:eq19}
\end{eqnarray}
The principal axis of the Dirac cone has an angle $\phi$ 
measured from the $k_y$ axis, where $-\pi/2 < \phi < \pi/2$.
When we denote the current and  electric field in this
   axis direction as $j_x'$ and $E_x'$, 
we obtain 
\begin{subequations}
\begin{eqnarray}
\begin{pmatrix}
  j_x'  \\
  j_y'
\end{pmatrix}  
=
\begin{pmatrix}
  \sigma_-  & 0  \\
  0  & \sigma_+
\end{pmatrix}  
\begin{pmatrix}
  E_x'  \\
  E_y'
\end{pmatrix} \; , 
 \label{eq:eq20a}
\end{eqnarray}
where 
\begin{eqnarray}
\begin{pmatrix}
  j_x'  \\
  j_y'
\end{pmatrix}  
=
\begin{pmatrix}
  \cos \phi & \sin \phi  \\
  - \sin \phi & \cos \phi
\end{pmatrix}  
\begin{pmatrix}
  j_x  \\
  j_y
\end{pmatrix} \; , 
 \label{eq:eq20b}
\end{eqnarray}
%
\begin{eqnarray}
\begin{pmatrix}
  E_x'  \\
  E_y'
\end{pmatrix}  
=
\begin{pmatrix}
  \cos \phi & \sin \phi  \\
  - \sin \phi & \cos \phi
\end{pmatrix}  
\begin{pmatrix}
  E_x  \\
  E_y
\end{pmatrix} \; . 
 \label{eq:eq20c}
\end{eqnarray}
\end{subequations}
The relations between ($\sigma_-, \sigma_+$, $\phi$) 
 and $(\sigma_x, \sigma_y , \sigma_{xy})$ are  
\begin{subequations}   
\begin{eqnarray}
\sigma_x& =& (\cos^2 \phi) \;  \sigma_- + (\sin^2 \phi) \; \sigma_+
 \; ,
    \label{eq:eq21aa}
          \\
\sigma_y &=&  (\sin^2 \phi) \; \sigma_- + (\cos^2 \phi) \; \sigma_+  
  \; , 
         \label{eq:21bb} 
    \\
\sigma_{xy} &=& \sin\phi \cos\phi\; (\sigma_- -  \sigma_+)
 \; , 
              \label{eq:eq21cc}
\end{eqnarray}
\end{subequations}  
or inversely,  
\begin{subequations}   
\begin{eqnarray}
\tan 2 \phi &=& \frac{2\sigma_{xy}}{\sigma_x-\sigma_y} \; ,
    \label{eq:eq21a}
          \\
 \sigma_{-} &=& 
 \frac{1}{2}[\sigma_x+\sigma_y - \sqrt{(\sigma_x-\sigma_y)^2+ 4\sigma_{xy}^2}] \; , 
            \nonumber \\
         \label{eq:21b} 
    \\
  \sigma_{+} & = &
     \frac{1}{2}[\sigma_x+\sigma_y + \sqrt{(\sigma_x-\sigma_y)^2+ 4\sigma_{xy}^2}] \;. 
       \nonumber \\
       \label{eq:eq21c}
\end{eqnarray}
\end{subequations}   
Note that $\sigma_{xy}$ is not a Hall conductivity and  $\sigma_{xy}=\sigma_{yx}$ holds 
as in Eq. (\ref{eq:eq19})
in the case of zero magnetic field.\cite{Kubo} 
$\sigma_{xy}$ is finite when $\sigma_- \ne \sigma_+$. 
The sign of $\phi$ is chosen such that   $\phi < 0$ for $\sigma_{xy} > 0$ and        $\phi > 0$ for $\sigma_{xy} < 0$,  where 
   $0 <|\phi| < \pi/4$ for $\sigma_y > \sigma_x$ and  
   $\pi/4  <|\phi| < \pi/2$ for $\sigma_x > \sigma_y$. 
 
\section{Results}
We calculate the conductivity   for the TB model with transfer energies 
      shown in  Fig.~\ref{fig1}. 
The direction of molecular stacking is given by the $y$ axis and  
 that perpendicular to the stacking is given by the $x$ axis. 
The NN transfer energies are examined for both 
 uniaxial pressure and hydrostatic pressure, while 
 those of NNN sites are added  only for hydrostatic pressure.\cite{Kino2006} 
In the following calculations,  the conductivity is normalized by 
  $e^2/\hbar$ and  the energy is scaled by eV.
 First, we  examine  the case of  uniaxial pressure ($P=6$ kbar) without  site potential and then  the  case of hydrostatic pressure with site potential 
to comprehend the similarity and dissimilarity.

\subsection{\ET\;  under uniaxial pressure}
 For the uniaxial pressure $P$ (kbar),
we take only  transfer energies with NN neighbor sites, 
 $t = a_1, \cdots, b_4$ (eV), which 
 are   estimated
 by the extended H\"uckel method and interpolation,   
\cite{Kondo2005,Katayama2006_JPSJ75} 
\begin{eqnarray}
t(P) = t(0) (1+K_t P)  \; , 
 \label{eq:eq20}
\end{eqnarray} 
 where $t(0)$ is given by 
$a_1(0), \cdots, b_4(0)$ = 
$ - 0.028$, $ - 0.048$, 0.020, 0.123, 0.140, 0.062, and 0.025,
 and the corresponding $K_t$ is written as 
 0.089, 0.167, $ -0.025$, 0, 0.011, and 0.032, respectively. 
Since the ZGS is obtained for $P > 3$ kbar, we choose  $P$ = 6 kbar as a typical pressure.

\begin{figure}
  \centering
\includegraphics[width=4cm]{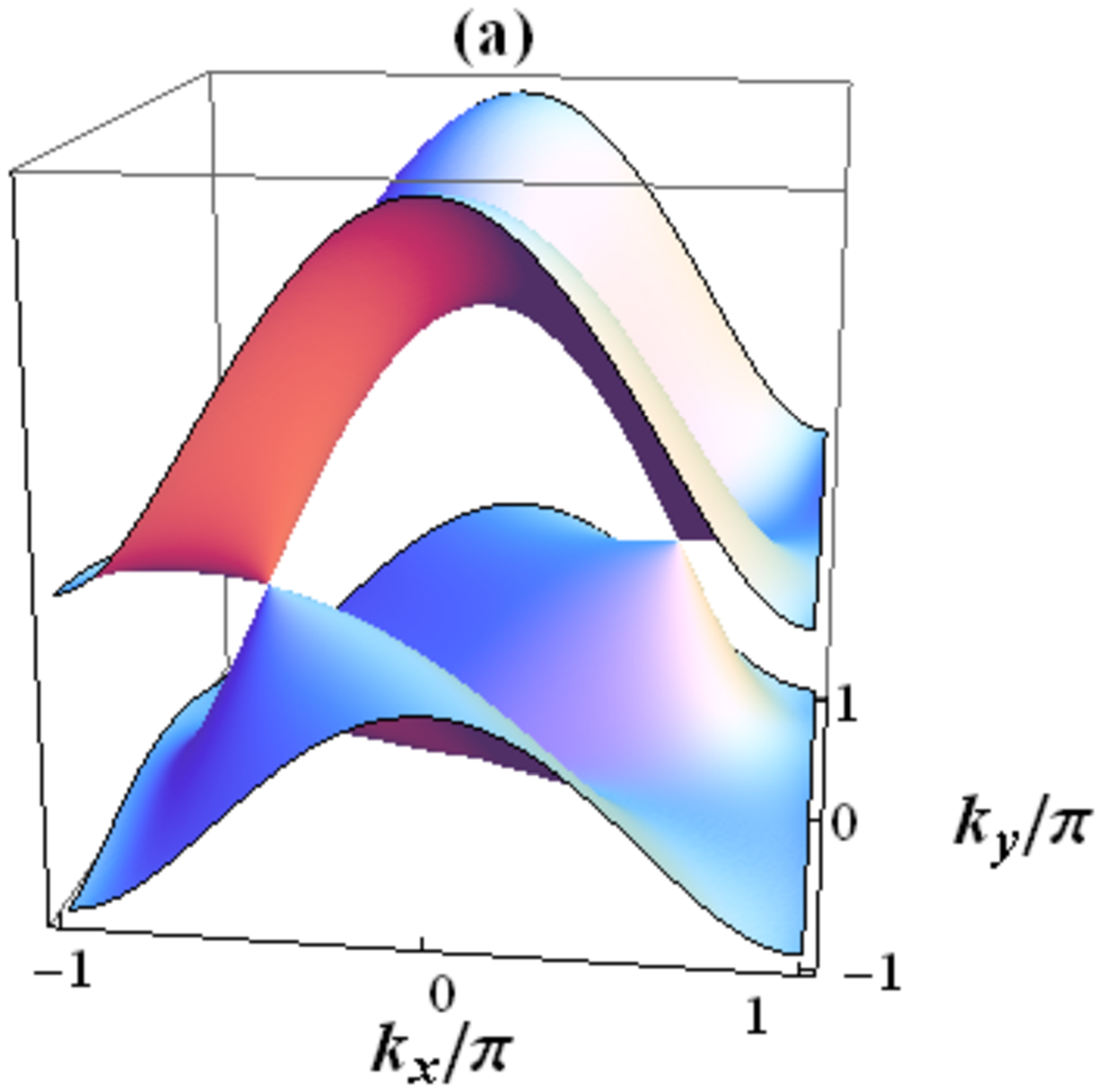}
\includegraphics[width=4cm]{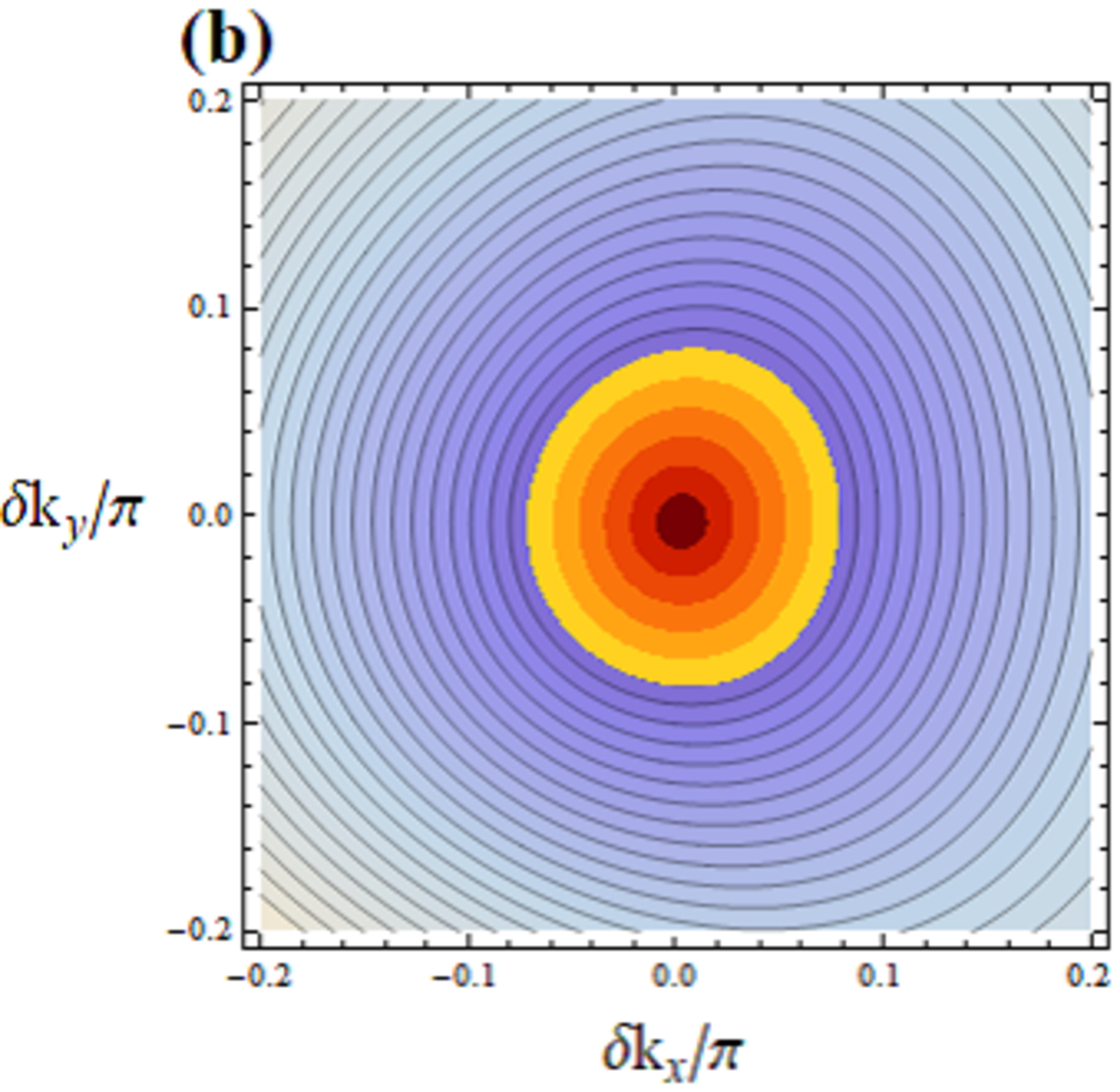} \\
\includegraphics[width=4cm]{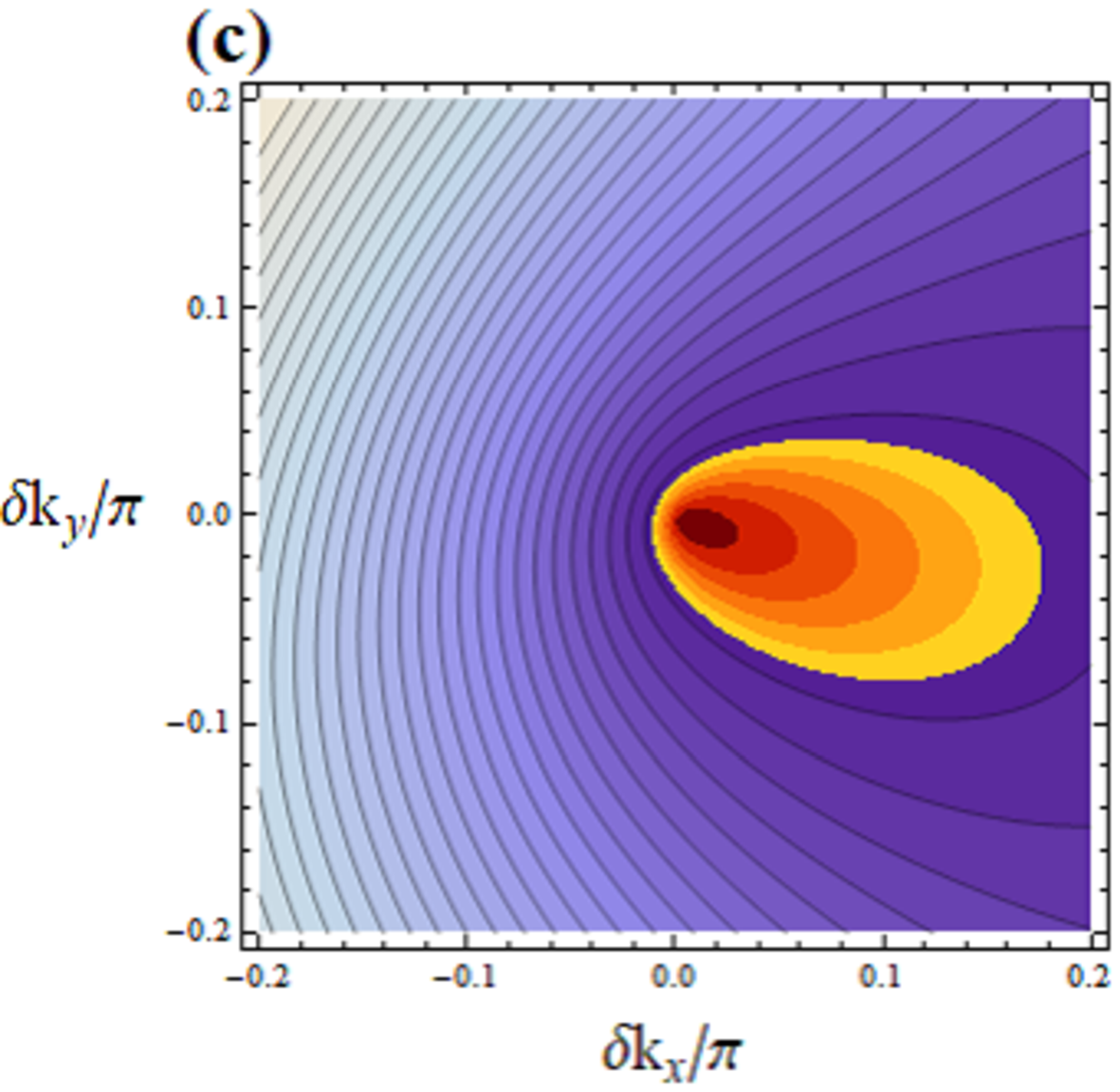}
\includegraphics[width=4cm]{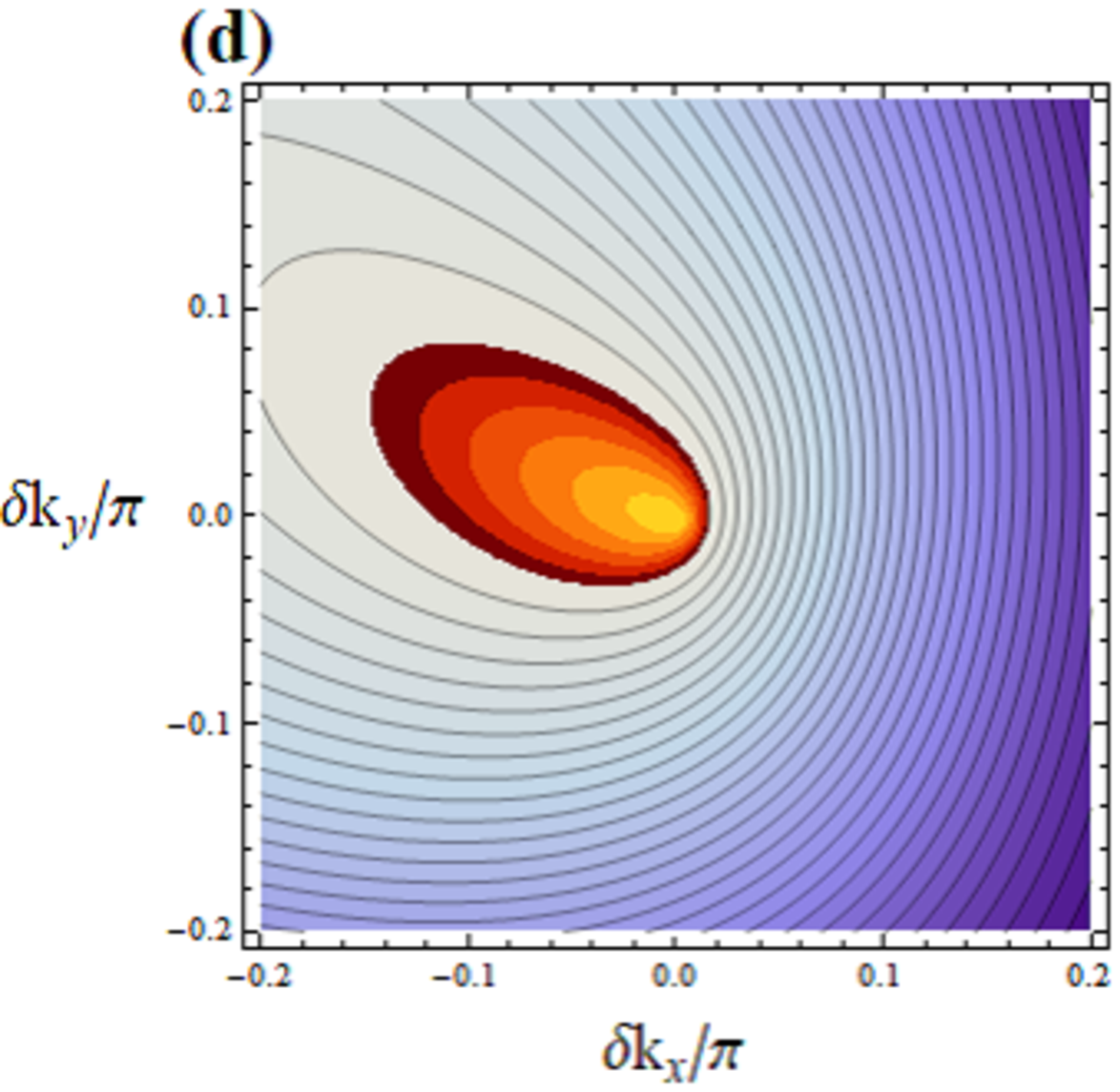} \\
\includegraphics[width=4cm]{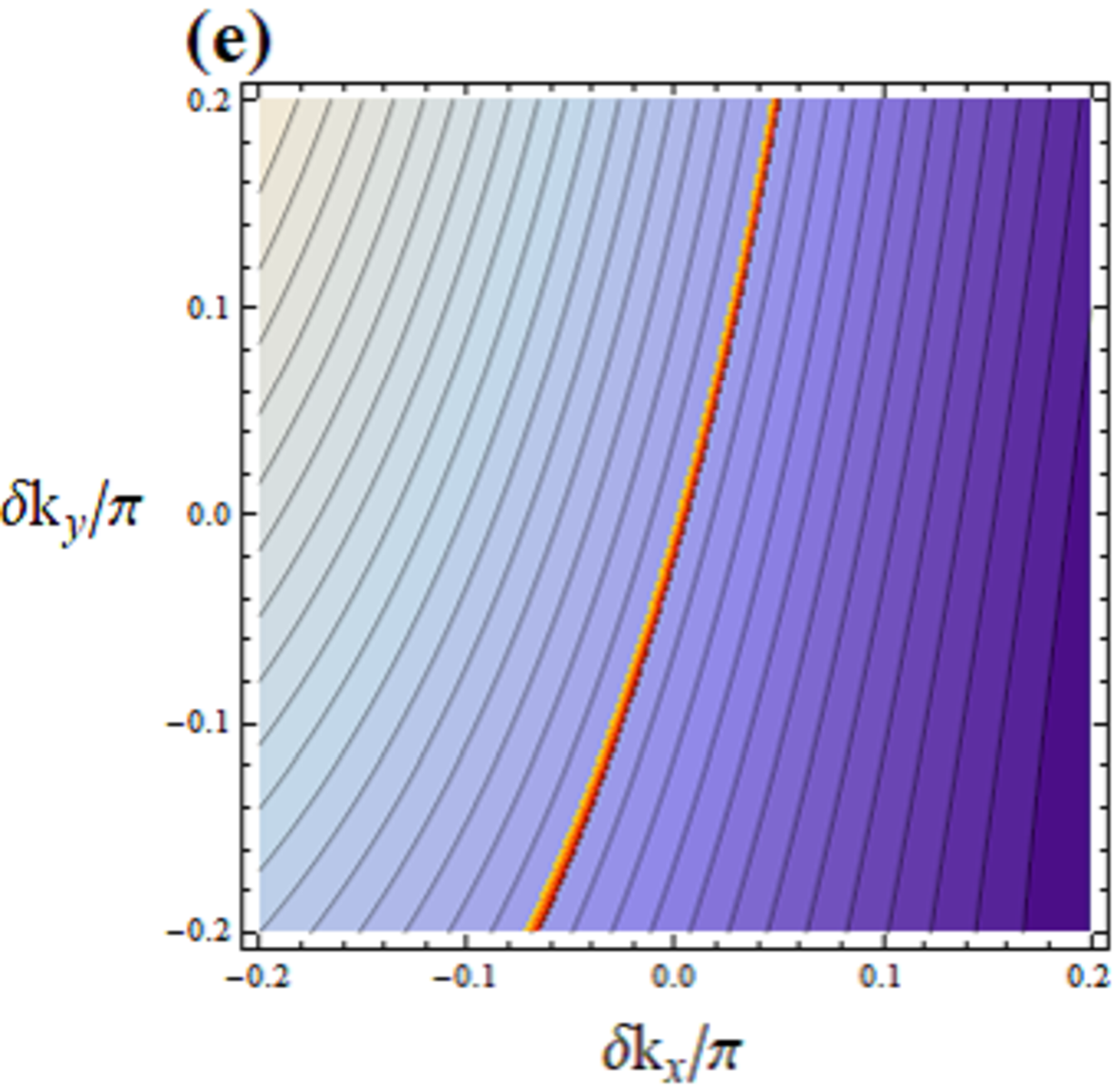} \\
     \caption{(Color online)
(a) Conduction and valence bands given by  
$E_1(\bk)$ (upper band) and $E_2(\bk)$ (lower band) 
at $P$= 6 kbar.
 Two bands contact at the Dirac points $\pm \bkD = \pm (0.57, 0.30)\pi$
  with an energy $\eD = \mu = 0.178$.
 (b) Contour plots of   $E_1(\bk) - E_2(\bk)$
 with the energy range [0, 0.106], where  
         $\delta \bk (= \bk - \bkD)$.
 The yellow line (outermost bright line) corresponds to 
$E_1(\bk) - E_2(\bk)$ =  0.03. 
 (c) Contour plots  of $E_1(\bk) - \eD$ with the range [0, 0.096], where  
   the Dirac point exists at (0,0). 
 The yellow line corresponds  to 
  $ E_1(\bk) - \eD$  = 0.005.  
 (d) Contour plots of $E_2(\bk) - \eD$ with the range [-0.076, 0]. 
  The Dirac point exists at (0,0). The outermost bright line corresponds to $ E_2(\bk) - \eD $ = -0.005. 
 (e) Contour plots of $E_1(\bk) + E_2(\bk) - 2\eD$ with the range [-0.062, 0.086]. 
    The bright color line denotes  $E_1(\bk) + E_2(\bk) - 2\eD = 0$.
  }
\label{fig2}
\end{figure}

 Figure \ref{fig2}(a) shows two bands of $E_1(\bk)$ and $E_2(\bk)$, 
 which 
 touch at  Dirac points $\pm \bkD = \pm (0.57, 0.30)\pi$ 
 with an energy $\eD$ = $\mu = 0.178$ corresponding to the three-quarter-filled band.
The ranges of the energy of the conduction and valence 
  bands $E_1(\bk)$ and $E_2(\bk)$  are  given by 
   $0 <E_1(\bk)-\eD < 0.19$ and $-0.11 < E_2(\bk)-\eD< 0$, respectively. 
Such  ZGS shows  the relation 
  $E_2(Y) < \eD < E_1(X) < E_1(M)$, where  $\Gamma$, X, Y, and M are TRIMs  given by  
 $\Gamma=(0,0)\pi$, $X=(1,0)\pi$,
 $Y=(0,1)\pi$, and $ M=(1,1)\pi$.
Figure \ref{fig2}(b) shows  
  contour plots of $ E_1(\bk)-E_2(\bk)$ 
as the function of 
 $\delta \bk = \bk -\bkD$ in a small region 
 around  $\bkD$. 
 The contour lines form almost isotropic circles, 
suggesting that 
 the velocity of the  Dirac cone is  isotropic.  
Figure \ref{fig2}(c) 
 shows $E_1(\bk) - \eD$. 
The Dirac point is located at 
 $(\delta k_x, \delta k_y)$ = (0,0).
 This contour  
suggests a tilted  Dirac cone and  
 shows a slight  deviation from the ellipse.
It is also found that 
 the cone shows  
 a slight rotation   clockwise from the  $k_x$ axis, which 
 plays a crucial role in the transport property, as shown later.
Figure \ref{fig2}(d) 
 shows $E_2(\bk) - \eD$. 
The Dirac point is located at (0,0).
 The contour of $E_2(\bk) - \eD$  
 also shows  a tilted  Dirac cone and a slight deviation from the ellipse. 
We define  a phase  $\phi_1 (<0)$ (
$\phi_2 $) as 
  a tilting angle of $E_1(\bk)$ ($E_2(\bk)$)  measured from the $k_x$  axis. 
Since $E_1(\bk)$ and $E_2(\bk)$ form a pair of Dirac cones, 
 $\phi_2 - \phi_1 = \pi$ for $\bk$ in the limit of the  Dirac point. 
 The deviation from the limiting value 
 increases   with increasing $|\delta \bkD|$. 
The tilting parameter, which is a ratio of the tilting velocity to 
 that of the Dirac cone,   is given by  $\eta \sim 0.8$. 
Figure \ref{fig2}(e) shows   contour plots of $E_1(\bk) + E_2(\bk) -2\eD$. 
 The  bright line, $E_1(\bk) + E_2(\bk) - 2\eD= 0$, 
 is perpendicular to the tilting direction of the cone.

\begin{figure}
  \centering
\includegraphics[width=7cm]{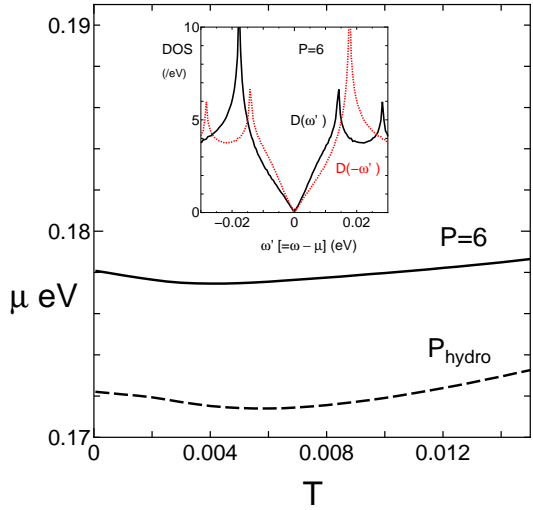}
     \caption{(Color online)
Chemical potential ($\mu$) as a function of temperature ($T$)   for $P$ =6  (solid line) 
 and $P = P_{\rm hydro}$ (dashed line).
It is found  that 
$\mu(T)$ for $P$=6 ($P = P_{\rm hydro}$) takes a minimum 
at  $T \sim 0.004$  ($T \sim 0.006$). 
The inset denotes the corresponding DOS as a function 
 of $\omega -  \mu=\omega'$ 
 for $P$ = 6, where D($-\omega$')(red dotted line) is compared with D($\omega$')
 showing the deviation from the linear dependence. 
}
\label{fig3}
\end{figure}

Figure \ref{fig3} shows the $T$  dependence 
 of the chemical potential $\mu(T)$ 
 for  $P$ = 6 (solid line) and a hydrostatic pressure 
   $P_{\rm hydro}$ (dashed line) examined in the next subsection, 
 where the corresponding DOS as a function of $\omega - \mu$ 
 is shown in the inset. 
 With increasing $T$,   $\mu$ varies slowly, suggesting that 
 the $T$ dependence of $\mu$ on $\sigma_\nu$ is negligibly small. 
In fact, the $T$ dependence of $\mu$  in Eq.~(\ref{eq:eq16a}) 
may be ignored for $\sigma_{\nu}(T)$ at low temperatures.
We verified that  $\mu(T)$ can be replaced by $\mu(0)$ in $\xi_{\g,\bk}$
 for $0 <T<0.015$ ($0<T<0.013$), which is the range of temperatures 
 of the following  numerical calculation of $\sigma_{\nu}$ under 
 the  pressure of $P$= 6 kbar ($P_{\rm hydro}$).
 The DOS close to the chemical potential  shows a clear deviation from 
the linear dependence with respect to $\omega - \mu$,  where $\mu$ is the chemical potential at $T$=0. 
 Such asymmetry of DOS is ascribed to the nonlinear spectrum of  
 the Dirac cone, where  $E_1(\bk)$  ($E_2(\bk)$)  is convex downward (upward) 
 on the tangent plane at the Dirac point. 
The slight decrease in the chemical potential   
 at low temperatures  comes from such asymmetry. 
 The increase in $\mu$ above  the minimum  occurs since 
 the van Hove singularity  at $E_2(Y)$ below the chemical potential  
has a large peak  compared with that  of  $E_1(X)$. 

\begin{figure}
  \centering
\includegraphics[width=7cm]{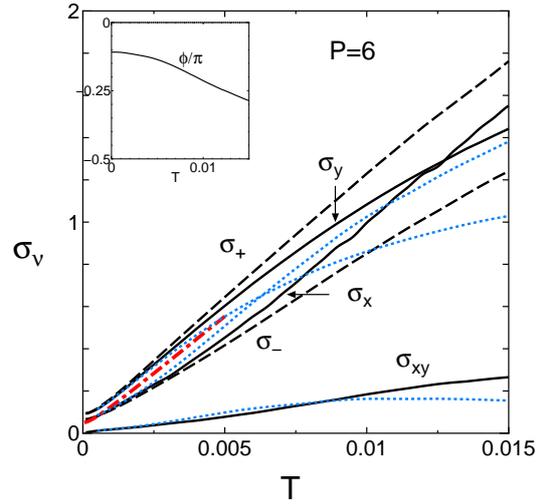}
     \caption{(Color online)
$T$ dependence of conductivity in the absence of  the e--p interaction 
 at $P$ = 6 with fixed $\Gamma$ = 0.0005, where 
 the solid lines denote $\sigma_x$, $\sigma_y$, and $\sigma_{xy}$ 
 and the dashed lines denote $\sigma_\pm$.
The dotted lines correspond to $\Gamma = 0.001$.
As a reference, 
 the conductivity of the isotropic Dirac cone without tilting, 
$\sigma_{\rm iso}$, 
 is shown   by the (red) dot-dashed line for $0< T <0.005$. 
 Principal values of $\sigma_{-}$  and $\sigma_{+}$
 are given by Eqs.~(\ref{eq:21b}) and (\ref{eq:eq21c}), respectively, 
 whereas $\phi$ is given by Eq.~(\ref{eq:eq21a}).
 The inset shows the phase $\phi$, which is  an angle of 
 the principal axis of $\sigma_-$  measured from the $k_x$ axis.  
}
\label{fig4}
\end{figure}

Figure \ref{fig4} shows the temperature dependence of 
conductivity of  $\sigma _\nu$ 
($\nu = x, y,xy, +, -$) without the e--p interaction, where  
   $P$ = 6 and  $\Gamma$ = 0.0005. 
Note that 
 the solid line was already obtained 
 in Ref.~\citen{Katayama2006_cond} but  $\sigma_{xy}$ was absent  
 since the energy band with the Dirac cone was found later.\cite{Katayama2006_JPSJ75}
We discuss $\sigma_{\nu}(T)$ by dividing temperatures  into two regions:
  (I) low temperatures with $\sigma_y$ being larger   than  $\sigma_x$ and  
  (II) high temperatures with $\sigma_x$ being larger than $\sigma_y$.

First, we examine  region (I), where the  quantum region is found,  
leading to the finite conductivity even at $T$ = 0. 
 In fact,  the conductivity in the zero  limit of $T$  is given 
 by  
$\sigma_x(0) \simeq 0.06$ and 
$\sigma_y(0) \simeq 0.09$,  which  
are compared with that of the universal value of 
 $1/2\pi^2 = 0.051$ obtained for the  simple Dirac cone.  
The slightly larger value in the present case 
 comes from the tilting of the Dirac cone.
\cite{Suzumura_JPSJ_2014}
An equality $\sigma_y > \sigma_x$
 at low temperatures 
is understood as follows.   
 Figure \ref{fig2}(b) shows  a nearly  isotropic velocity of the Dirac cone, 
 whereas  Figs.~\ref{fig2}(c) and \ref{fig2}(d) present 
   a large tilting of the Dirac cone  almost  along the $k_x$ direction. 
Our previous calculation of a  tilted Dirac cone at $T$ = 0
 shows that the conductivity becomes maximum for the direction 
 perpendicular to  a tilting axis.~\cite{Suzumura_JPSJ_2014}  
This finding  explains   $\sigma_y > \sigma_x$ 
   in the present case. 
 Such inequality still holds at finite temperatures  
 above $T > \Gamma$.  
As a reference, 
 the conductivity of the isotropic Dirac cone without tilting, 
$\sigma_{\rm iso}$, 
 is shown   by the dot-dashed line for $0< T <0.005$. 
With increasing $T$,  $\sigma_{\rm iso}$, 
which is slightly lower than $\sigma_x$ at $T=0$,
 becomes larger than $\sigma_x$.
 Thus, the effect of tilting is found in  region (I), 
 where  $\sigma_y > \sigma_{\rm iso} > \sigma_x$.  
 By comparing the solid line ($\Gamma$ = 0.0005) 
 with the dotted line ($\Gamma$ = 0.001), we found that  
  the effect of tilting is reduced for large $\Gamma$.

Next, we examine  region (II), where 
$\sigma_x > \sigma_y$ is found for  $T > 0.012$. 
 In  region (II) the effect of the tilting is neglected but 
 the difference between  $\sigma_x$ and $\sigma_y$ still exists owing to 
 the anisotropy of the velocity of the cone.  
Since  the DOS is averaged 
    owing to a factor $( - \partial f(\ep) /\partial \ep)$ 
 in Eq.~(\ref{eq:sigma}) having  a broad distribution 
   with a width $\sim T$,
this resembles  a classical  conductivity
 with  a finite chemical potential being $\sim T$.
Thus, we  find $\sigma_x  > \sigma_y$, 
 since  the average velocity 
 is larger for the $x$-direction than for the $y$-direction  
  because the  transfer energies are larger for the $x$-direction than 
 for the  $y$-direction.
Such  crossover occurs at lower $T$ for larger $\Gamma$. 
We also note that such crossover occurs at higher $T$
  for $P$ = 8 (not shown here). 

The dashed lines in Fig.~\ref{fig4} show the $T$ dependence of 
 the principal value $\sigma_{\pm}$.  
In the limit of $T=0$, 
 the principal conductivity $\sigma_+$, which gives the maximum of the 
 conductivity, becomes $\simeq \sigma_y$ 
  and is 
 perpendicular to  the tilted Dirac cone.
 Note that this direction is almost parallel to the line 
of $E_1(\bk)+E_2(\bk)=0$ (see Fig.~\ref{fig2}(e)).
 For $T \rightarrow 0$, we obtain 
 $\sigma_+(0) \simeq 0.092$ and  $\sigma_-(0) \simeq 0.0601$, 
  which give 
   $\sigma_+(0)/\sigma_0 \simeq 1.82$ and  
    $ \sigma_-(0)/\sigma_0 \simeq 1.32$ 
   with  $\sigma_0=1/(2\pi^2)$. 
 These results are consistent  with the analytical results of 
 the  tilted Dirac cone,  where 
  $\sigma_+/\sigma_0$  = 1.84 and
  $\sigma_-/\sigma_0$  =1.19 for $\eta$ =084.~\cite{Suzumura_JPSJ_2014} 
The behavior of $\sigma_{\pm}$  being linear in $T$ 
 resembles that of DOS around $\omega=\mu$ in the inset of 
 Fig.~\ref{fig3}, which is found for $|\omega -\mu| < 0.015$.  
The inset shows the  $T$ dependence of $\phi$, where   
  $\phi < 0 $  owing to  $\sigma_{xy} > 0$  and 
$\sigma_y > \sigma_x$  at low temperatures, 
 as seen from Eq.~(\ref{eq:eq21a}). 
 With increasing $T$, $\phi$ decreases and becomes smaller than $-\pi/4$ 
 at a temperature corresponding to  
 $\sigma_{x}=\sigma_y$, where the axis close to  $\sigma_-$ 
 changes  from the $k_x$ axis  to the $k_y$ axis.  
Thus, the principal axes  
 rotate clockwise with an angle $\phi (<0)$
 (see Eq.~(\ref{eq:eq20c})). 

\begin{figure}
  \centering
\includegraphics[width=7cm]{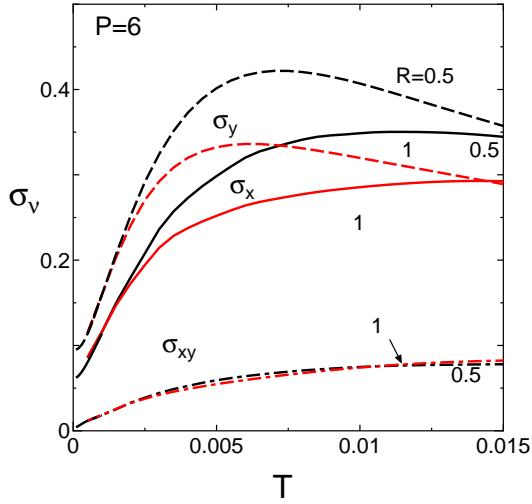} 
     \caption{(Color online)
Conductivity $\sigma_{\nu}$ ($\nu = x, y$, and $xy$)
in the presence of the e--p interaction
with $P$ = 6 and $\Gamma$ = 0.0005. 
 The solid, dashed, and dot-dashed lines 
 denote  
 $\sigma_x$, $\sigma_y$, and 
 $\sigma_{xy}$, respectively, where 
  $R$ denotes a normalized e--p coupling constant defined  by 
   the ratio  $\lambda/\lambda_0$ (Eq.~(\ref{eq:eq16b})). 
  $R$ = 1 corresponds to 
  $\lambda =0.1$. 
At high $T$, 
 $\sigma_x$ shows  almost constant behavior and 
 $\sigma_y$ shows a broad maximum.
}
\label{fig5}
\end{figure}

\begin{figure}
  \centering
\includegraphics[width=7cm]{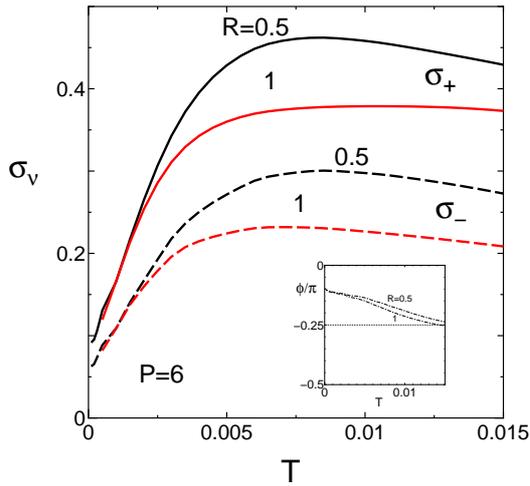} 
     \caption{(Color online)
Conductivity $\sigma_\pm$  with $P$ = 6 and $\Gamma$ = 0.0005
in the presence of the e--p interaction.
The solid and dashed lines correspond to 
 $\sigma_+$ and $\sigma_-$, respectively, 
 which show  a broad maximum.
The inset denotes the corresponding $\phi$.
}
\label{fig6}
\end{figure}

 As shown in Fig.~\ref{fig4} ( $P$=6),
  $\sigma_{\nu}$ increases monotonically  as a function of $T$, 
 which is different from the experiment showing 
   nearly constant behavior at high temperatures.~\cite{Kajita_JPSJ2014}  
Such an exotic $T$ dependence of $\sigma_{\nu}$ is examined next  
  by taking account of  the e--p interaction, 
      which is expected to reduce $\sigma_{\nu}$. 
By using  Eq.~(\ref{eq:damping}) and (\ref{eq:eq16a}),
   we calculate $\sigma_{\nu}$  of Eq.~(\ref{eq:sigma}), 
     where  $\Gamma$ in the absence of the e--p interaction 
      is replaced by 
        $\Gamma_{\g} (= \Gamma + \Gamma_{\rm ph}^{\g})$. 
Owing to the $T$ dependence of   $\Gamma_{\rm ph}^{\g}$, 
   $\Gamma$ is dominant at low $T$  
     whereas $\Gamma_{\rm ph}^{\g}$ is dominant at high $T$.
Note that  such  crossover with increasing $T$ depends on $R$. 

Figure \ref{fig5} shows the  $T$ dependence of 
    $\sigma_{\nu}$ ($\nu = x, y,$ and $xy$) 
       in the presence of the e--p interaction 
  with some choices of $R$. 
 The effect of $R \not= 0$ appears when   
 $\sigma_{\nu}$ deviates from the $T$-linear behavior. 
  Compared with  $\sigma_{\nu}$ with $R$ = 0  (Fig.~\ref{fig4}),
 $\sigma_{\nu}$  is reduced  noticeably. 
  At temperatures around $T \sim 0.015$,  $\sigma_x$ 
    becomes nearly constant,  
      whereas  $\sigma_y$  takes  a broad maximum at lower temperatures. 
The quantity  $\sigma_y -\sigma_x$ 
  decreases noticeably  at high temperatures,
 suggesting that the  effect of the tilting of the Dirac cone 
 decreases  with increasing temperature.
The $R$ dependence on $\sigma_{xy}$ is small.
The crossover temperature corresponding to  
$\sigma_x = \sigma_y$ decreases with increasing $R$,
 since the reduction of $\sigma_y$ is larger than 
 that of $\sigma_x$.
The nearly constant behavior of $\sigma_x$ is understood as follows.
With increasing $T$,   
$\sigma_{\nu}$ without the e--p interaction ($R = 0$) 
  increases linearly owing  to 
 the DOS obtained from the Dirac cone.
Such a linear increase is suppressed  for $R \not= 0$, since   
 the noticeable effect of the acoustic phonon  emerges 
 at finite temperatures.
In fact,  the electron is scattered by both normal impurity ($\Gamma$) 
 and the e--p interaction ($\Gamma_{\rm ph}^{\gamma}$), and 
  the  latter becomes dominant  at high temperatures 
    as seen from  Eq.~(\ref{eq:eq16a}).
 However, compared with a case of the conventional  
 metal with a Fermi surface,
  the effect of the  e--p scattering  
 in the case of the  Dirac cone close to the  three-quarter-filled  band 
 is  strongly  reduced 
 owing  to a constraint
  by the energy-momentum conservation.\cite{Suzumura_PRB_2018} 
   This could possibly give rise to a nearly constant behavior or  a broad maximum in $\sigma_{\nu}$,
      owing  to 
  a competition between the enhancement by  DOS of the Dirac cone and     
 the suppression by the e--p interaction. 

 Figure \ref{fig6} shows the $T$ dependence of the principal values 
   $\sigma_{\pm}$ corresponding to  Fig.~\ref{fig5}. 
Since $\sigma_+$ and $\sigma_-$ give the upper and lower bounds 
 of the conductivity, these quantities are convenient 
 for comprehending  the results of experiments 
 even when  the precise relationship  
 between the crystal axis and the direction of 
the applied electric field is unknown. 
In the inset, the  angle between the $k_x$ axis and the tilting axis 
 is shown, where 
$\phi < 0 $ for $\sigma_{xy} > 0$ and $\sigma_y > \sigma_x$.
The angle $|\phi|$ increases and exceeds $\pi/4$ at a certain 
 temperature with $\sigma_x=\sigma_y$, implying that 
  the axis of $\sigma_-$ becomes 
   closer to the $k_y$ axis at high temperatures.
Such a behavior can be understood    
    on the basis of 
      a simplified model,~\cite{Suzumura_PRB_2018} where 
 $\sigma_+ \sim \sigma_- \sim \sigma$.
Note that 
 $\Gamma_{\rm ph}^\g$ is obtained in  Eq.~(\ref{eq:eq16a})
  and $\sigma  \simeq a_{\nu}'10^3T /\Gamma $ with  $a_{\nu}'= o(0.1)$
without e--p interaction. 
By taking $\Gamma$ replaced by $\Gamma + \Gamma_{\rm ph}^\g$ and 
employing an idea   $<|\xi_{\gamma,\bk}|> \sim T$ 
 with $<>$ being  an average value 
 in the summation of Eq.~(\ref{eq:self_energy}), we obtain  
\begin{eqnarray}
\sigma \simeq \frac{a_{\nu}'10^3T}{1 + C_0RT^2/\Gamma} \; ,
  \label{eq:eq22b}
\end{eqnarray}
with $C_0$ = 12.5 and $\Gamma$ = 0.0005. 
  From Eq.~(\ref{eq:eq22b}), it is found 
 that 
 a maximum of $\sigma$ as a function of $T$ is obtained  
 by a  competition of the  DOS (the numerator)
  and the e--p interaction (the denominator). 
 Equation (\ref{eq:eq22b}) suggests 
that $\sigma$ decreases  with increasing $R$. 

\subsection{\ET\;  under hydrostatic pressures}
We examine Dirac electrons under hydrostatic pressure 
 using the TB model  with 
 NN ($a_1, \dots, b_4$) 
 and  NNN ($a_{1d}, \cdots, a_{4d}$) 
transfer energies given by\cite{Kino2006} 
     $a_1 = -0.0267$,
      $a_2 = -0.0511$,
      $a_3=0.0323$,
      $b_1 =0.1241$,
      $b_2 =0.1296$,
      $b_3 =0.0513$,
      $b_4 =0.0152$,
      $a_{1d} =0.0119$,
      $a_{3d} =0.0046$, and 
      $a_{4d} =0.0060$. 
  The site potentials given by Eqs.~(\ref{eq:VB}) and (\ref{eq:VC})~\cite{Katayama_EPJ} are  estimated as  $\tV_B =0.0511$,  $\tV_C = 0.0032$ (Appendix A). 

\begin{figure}
  \centering
\includegraphics[width=4cm]{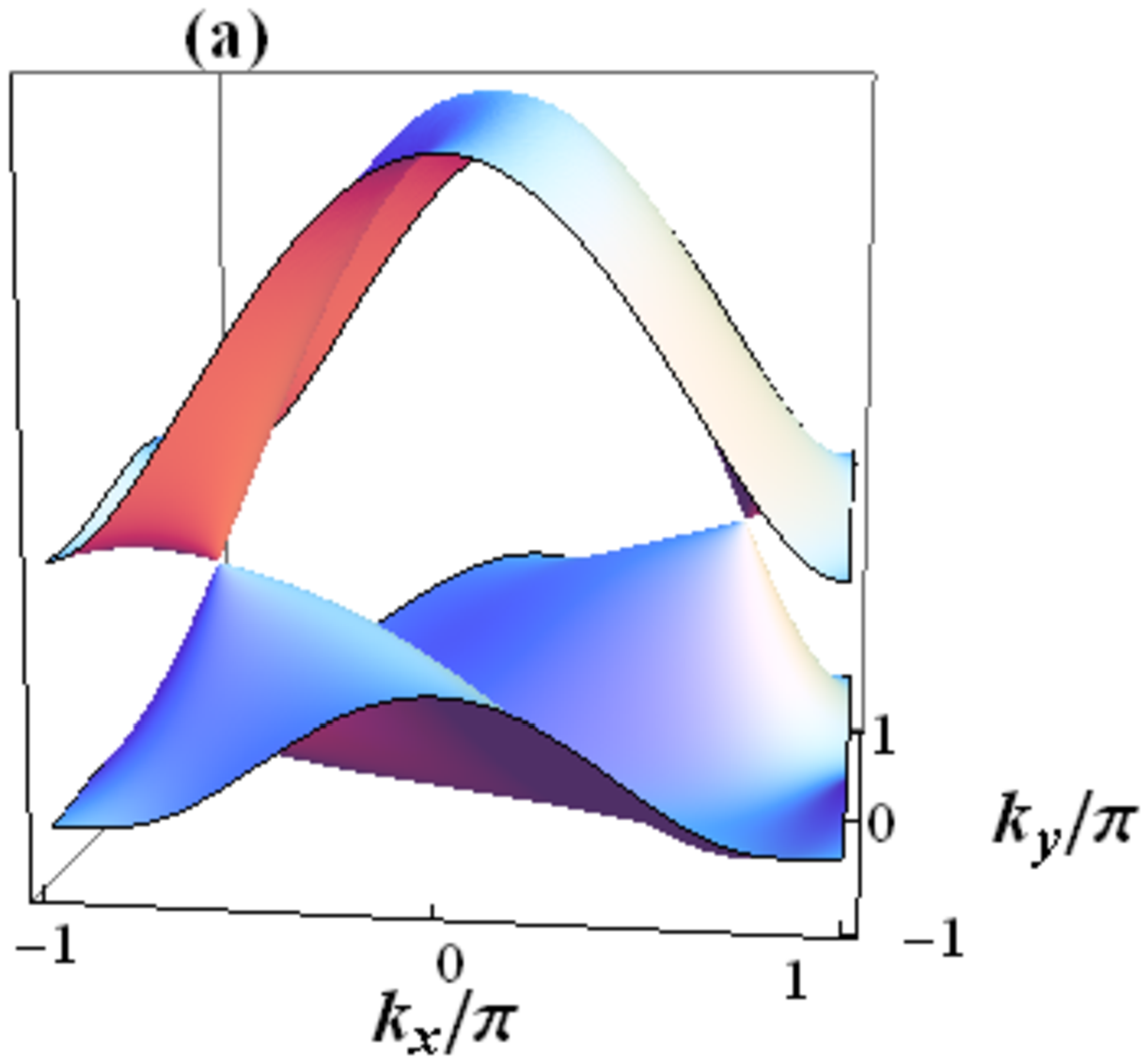}
\includegraphics[width=4cm]{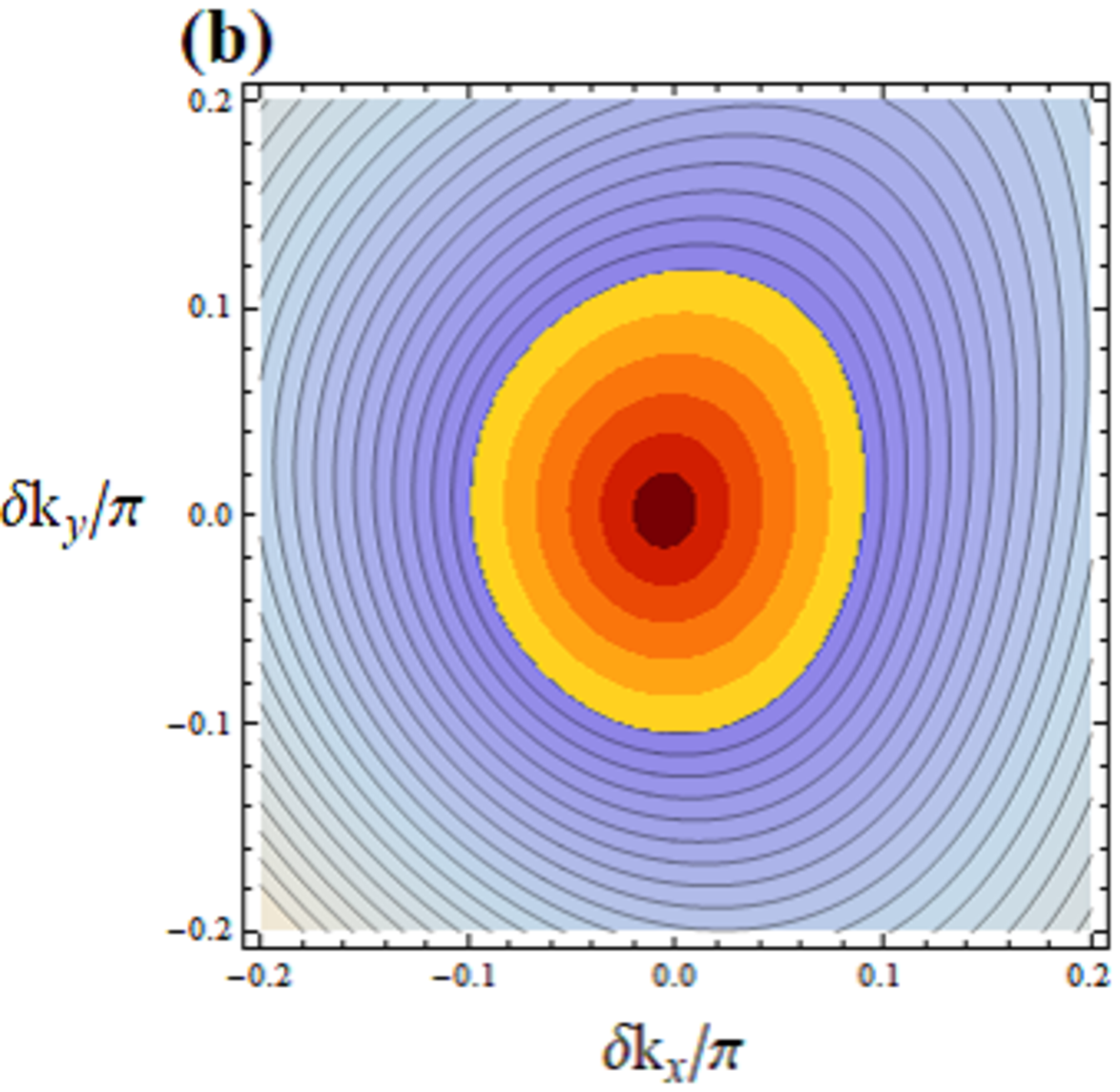} \\
\includegraphics[width=4cm]{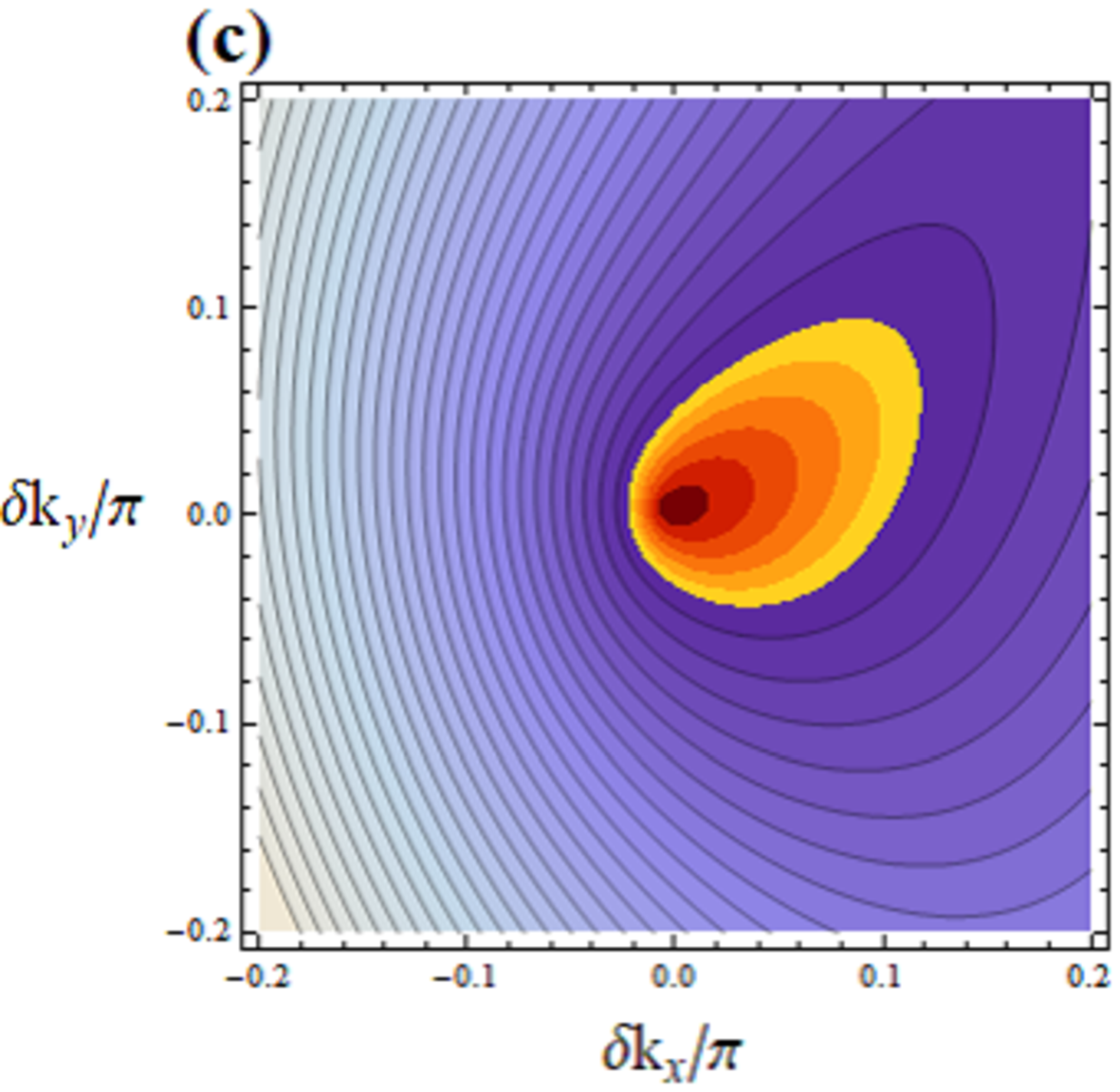}
\includegraphics[width=4cm]{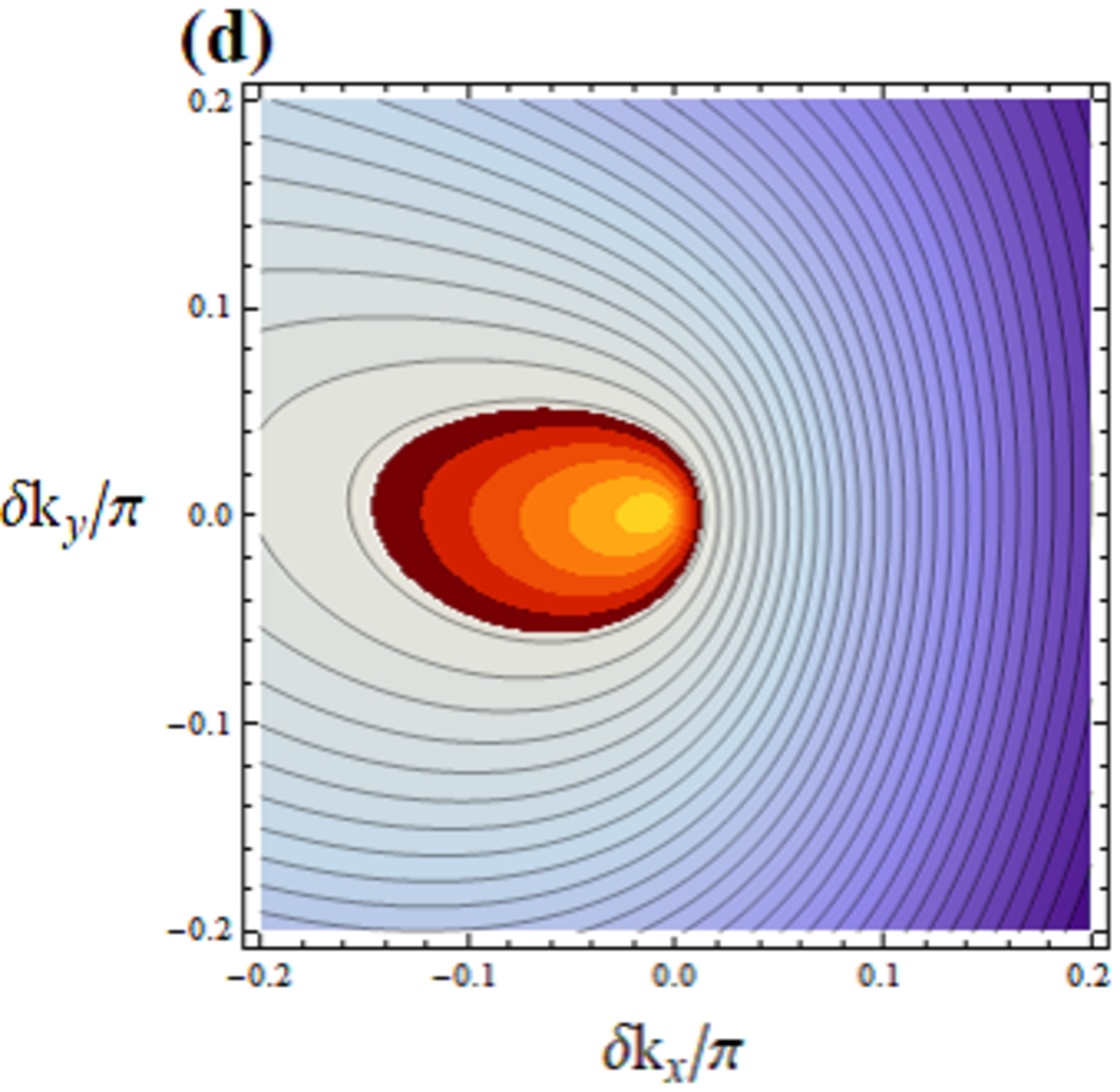} \\
\includegraphics[width=4cm]{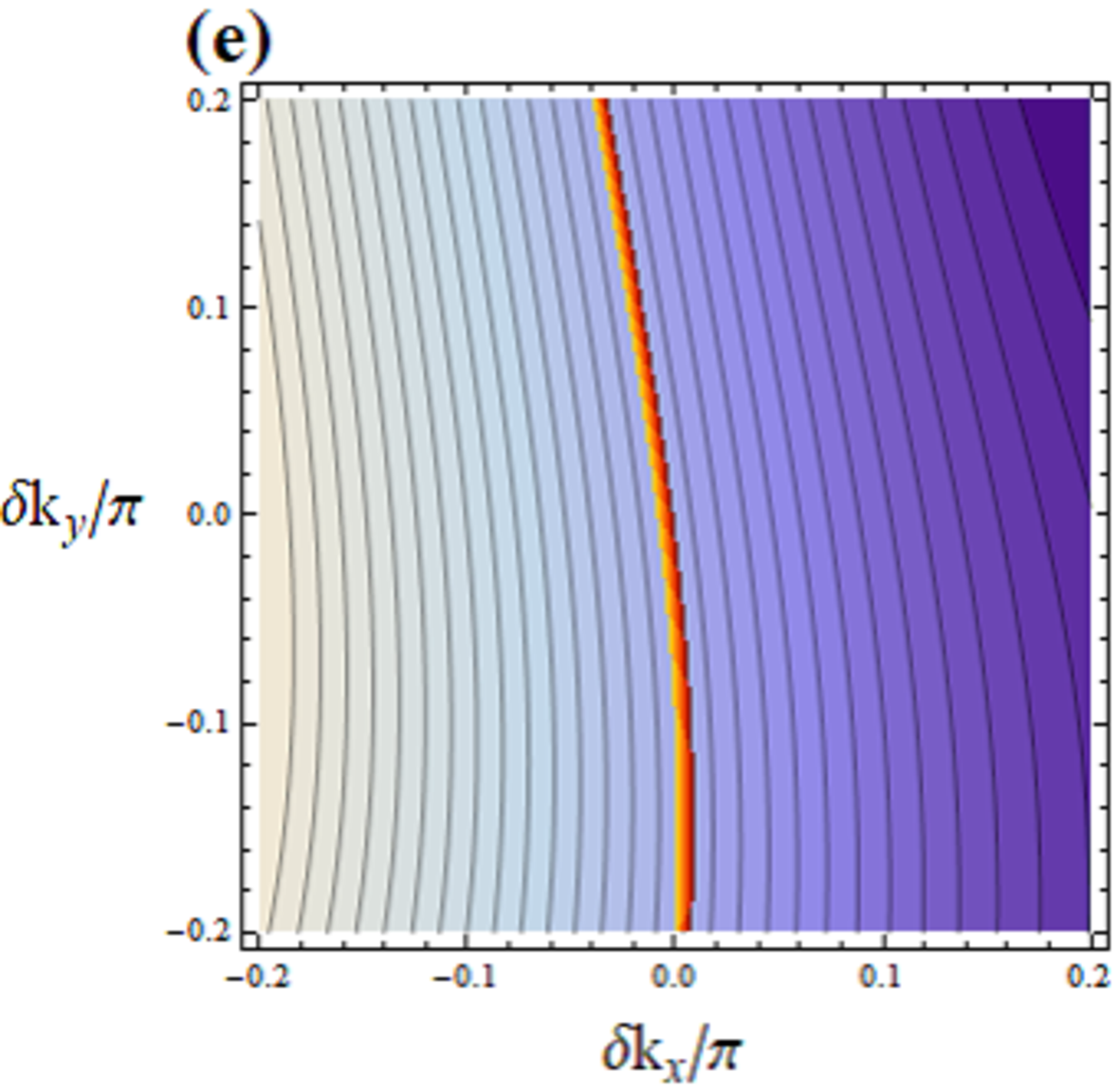} \\
     \caption{(Color online)
 Energy bands   of $E_1(\bk)$ and $E_2(\bk)$ under hydrostatic pressure,
  where the Dirac points are given by 
$\pm \bkD = \pm (0.69, 0.44)\pi$ and $\mu = \eD$ =0.172. 
(a) Conduction and valence  bands given  by  
 $E_1(\bk)$ (upper band) and $E_2(\bk)$ (lower band). (b) Contour plots of $E_1(\bk)- E_2(\bk)$ as a function of 
 $\delta \bk = \bk - \bkD$ with the energy range [0, 0.09]. 
 The yellow line (outermost bright line) corresponds to 
        $E_1(\bk)- E_2(\bk)$ =  0.03. 
(c) Contour plots of   $E_1(\bk)-\eD$   
   as a function of  $\delta \bk$ with the range [0, 0.066]. 
 The yellow line  corresponds to 
  $ E_1(\bk) -\eD$ =  0.005. 
(d) Contour plots of  $E_2(\bk)-\eD$  
   as a function of  $\delta \bk$ with the range [-0.056, 0].
    The outermost  colored region corresponds to 
  $ E_1(\bk) -\eD = - 0.005$.  
(e) Contour plots of $E_1(\bk) + E_2(\bk) - 2\eD$ with the range [-0.045, 0.045]. 
    The bright color line denotes  $E_1(\bk) + E_2(\bk) - 2\eD$ = 0.
}
\label{fig7}
\end{figure}

Band energies $E_j(\bk)$ under hydrostatic pressure, which     
  have Dirac cones similar to Fig.~\ref{fig2},  are as follows.  
Figure \ref{fig7}(a) shows the conduction and valence bands 
  ($0 <E_1(\bk)-\eD < 0.15$ and $-0.09 < E_2(\bk)-\eD < 0$) 
 in the first Brillouin zone,  
   which  touch at the Dirac points 
       $\pm \bkD = \pm (0.69, 0.44)\pi$ with $\eD$ =0.172. 
Compared with Fig.~\ref{fig2}(a), the band width of both 
$E_1(\bk)$ and $E_2(\bk)$ is slightly large and the Dirac points 
 move away from the $\Gamma$ point.
In Fig.~\ref{fig7}(b),  the energy difference 
 between $E_1$ and $E_2$ is shown. 
 The contour lines form ellipsoids, suggesting that  
the anisotropy of the velocity of the  Dirac cone is larger than that of 
Fig.~\ref{fig2}(b), e.g.,  
  an  ellipse with the ratio of the major to 
 minor axes being $\simeq 1.2$. 
  Figure \ref{fig7}(c) shows that  
 the  tilting angle $\phi (>0)$  of $E_1(\bk)-\eD$  
 measured from the $k_x$ axis 
 has the sign opposite to that of Fig.~\ref{fig2}(c).
  As shown later,  such difference in the sign results in the difference 
 in the current direction, i.e., the rotation  with respect to the applied 
   electric field of $E_x$ or $E_y$.  
Figure \ref{fig7}(d) shows $E_2(\bk)-\eD$ forming a pair of Dirac cones 
 with that of Fig.~\ref{fig7}(c), where  the deviation of the tilting axis from the $k_x$ axis 
 is   opposite to that of Fig.~\ref{fig2}(d).
The tilting direction is understood from  Fig.~\ref{fig7}(e), where  
the energy of the Dirac point is given by 
 $E_1(\bkD)-\eD =  E_2(\bkD)-\eD=0$ and 
  the tilting direction of the cone is almost perpendicular to the  
zero line  [$E_1(\bk) +  E_2(\bk) -2\eD=0$].
By comparing   Figs.~\ref{fig2} with  Figs.~\ref{fig7}, 
we find 
the opposite  rotation of the tilted cone.  

\begin{figure}
  \centering
\includegraphics[width=7cm]{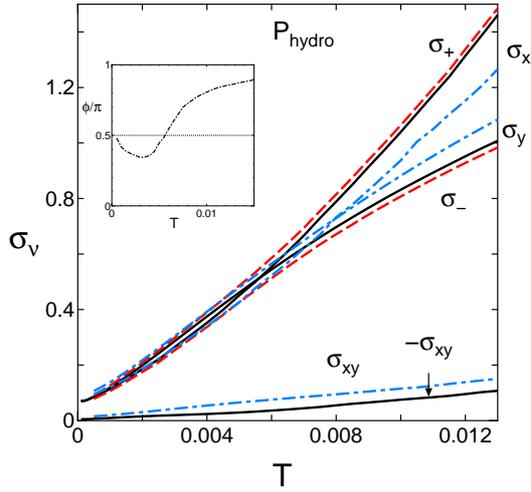}
     \caption{(Color online)
Conductivity $\sigma_{\nu}$ 
  with $\nu = x, y$, and $xy$ (solid line), 
 and $\nu = \pm$ (dashed line) under hydrostatic pressure.  
Note that $\sigma_{xy} < 0$ and $\phi/\pi (>0)$ (inset) 
from Eq.~(\ref{eq:eq21a}).
The dot-dashed line denotes $\sigma_\nu$ without the site potential, 
 where  $\mu (0)$ = 0.168
 }
\label{fig8}
\end{figure}

Now we examine the electric conductivity.
The parameters for  $P_{\rm hydro}$ are taken as  
 $\Gamma$ = 0.0005 and $\tV_B$= 0.0511, $\tV_C$ = 0.0032 for the site 
 potentials, which are obtained from Eqs.~(\ref{eq:VB}) and (\ref{eq:VC}). 
Figure \ref{fig8} shows 
  the conductivity $\sigma_{\nu}$ without the e--p interaction, 
 where the solid line (dashed line)  denotes
       $\sigma_x$,   $\sigma_y$, and  $\sigma_{xy}$ ($\sigma_{\pm}$).
Compared with the case of the uniaxial pressure (Fig.~\ref{fig4}),
 the sign with  $\sigma_{xy} < 0$  is  opposite,  
 as also seen from a comparison 
   of Figs.~\ref{fig7}(e) and  \ref{fig2}(e),  showing that 
   the angle of  the zero line $E_1+ E_2 -2\eD=0$ from the $k_y$ axis   is opposite.
This difference originates from 
the transfer energies.
A noticeable difference between  
 hydrostatic pressure and  uniaxial pressure 
  exists  in   $\sigma_{\nu}$  at low temperatures, 
  where $\sigma_y > \sigma_x$ in Fig.~\ref{fig4} and  
     $\sigma_y \simeq \sigma_x$ in Fig.~\ref{fig8}.
The origin of  $\sigma_y \simeq \sigma_x$ is understood as follows from 
Figs.~\ref{fig7}(b)--\ref{fig7}(d). 
Figure \ref{fig7}(b) shows an anisotropy of  the contour
 implying that the ratio of the  maximum to  minimum velocities  
 of the Dirac cone is $\simeq 1.2$, which results in 
 the ratio of the maximum and minimum conductivities being 
 $\simeq 1.4$.~\cite{Suzumura_JPSJ_2014} 
On the other hand,  the tilted cone  
 gives $\sigma_y / \sigma_x \simeq 1.4$ for the tilting parameter 
 $\eta \sim 0.9$.~\cite{Suzumura_JPSJ_2014} The competition of these two ingredients provides an almost 
  isotropic conductivity at low temperatures. 
The conductivity at high temperatures 
 is given  by the classical region, where 
 the effect of the tilting can be neglected  
 and  the  larger  transfer energies along  the $x$ direction 
  gives a larger velocity for the $k_x$ axis as seen from Fig.~\ref{fig7}(b).

Figure \ref{fig8} also shows the principal value $\sigma_{\pm}$ 
 of the conductivity, where 
 $\phi (>0)$  in the inset 
   denotes an angle between the $\sigma_-$ axis and the $k_x$ axis. 
With increasing $T$,  the difference between 
 $\sigma_+$ and $\sigma_-$  increases but is smaller 
 than  that shown in  Fig.~\ref{fig4}.
This comes from a difference at low temperatures, where 
$\sigma_y - \sigma_x$ for the uniaxial pressure is larger than 
that for the hydrostatic pressure.

We discuss   $\sigma_{\nu}$ with transfer energies of hydrostatic pressure 
 but without site potentials, 
which  is shown by the dot-dashed line in Fig.~\ref{fig8}. 
In this case,  $\sigma_y \simeq \sigma_x$ but $\sigma_{xy} > 0$.
This comes from the  fact that the ellipse corresponding to Fig.~\ref{fig7}(b)
 is slightly rotated anticlockwise (not shown here).
Thus, it turns out that  $\sigma_y \simeq \sigma_x$ comes from 
 transfer energies of hydrostatic pressure and that 
$\sigma_{xy} > 0$ is attributable to the absence of the interaction.

\begin{figure}
  \centering
\includegraphics[width=7.8cm]{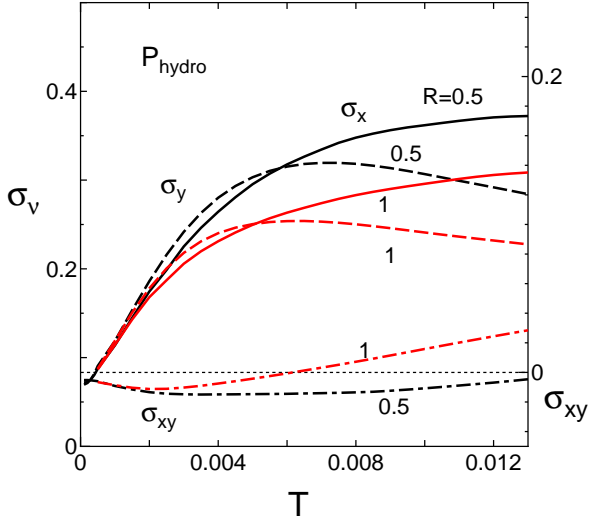}
     \caption{(Color online)
Conductivity 
$\sigma_{\nu}$ ($\nu$ = $x, y,$ and $xy$) under hydrostatic pressure 
 in the presence of the e--p interaction
with  choices of  $R$ = 0.5 and 1. 
}
\label{fig9}
\end{figure}

Now, we examine the effect of the e--p scattering 
on the $T$ dependence of conductivity with some choices of $R$,  Eq.~(\ref{eq:eq16b}), where  
the effect of the site potential is included in  the energy band. 
Figure \ref{fig9} shows $\sigma_{\nu}$ 
 with  the fixed $R$ = 0.5 and 1, 
 in which   
 sufficient  suppression of $\sigma_\nu$ by $R$ is seen  compared with   
 that of Fig.~\ref{fig8} ($R$ = 0). 
The  crossover temperature, where 
$\sigma_y > \sigma_x$  at low temperatures and    
 $\sigma_x > \sigma_y$   at high temperatures, 
 decreases with increasing $R$.  
The increase toward a  constant  behavior at   
 high $T$ is seen for $\sigma_x$,  
 whereas  a broad maximum is found for  $\sigma_y$. 
   $\sigma_{xy}$ as a function of $T$ takes a minimum  followed by 
the change of the sign for large R. 
 Compared with Fig.~\ref{fig4}, the crossover temperature of 
$\sigma_y = \sigma_x$
is lower for the hydrostatic pressure, where  
  $\sigma_y$ is sufficiently  reduced and  $\sigma_x$ is less reduced.

\section{Summary and Discussion}
We calculated the $T$ dependence of conductivity $\sigma_{\nu}$ 
($\nu = x$ and $y$)  of Dirac electrons in the organic conductor \ET\; under both uniaxial and hydrostatic  pressures 
to find that 
the anisotropic conductivity exhibits a crossover  from  region (I) 
including
the quantum regime at low temperatures to  region (II) 
 showing  the classical regime at high temperatures.
The former with $\sigma_y > \sigma_x$ 
 comes from the tilted Dirac cone, 
whereas   the latter with $\sigma_x > \sigma_y$ 
 originates from the anisotropy of the velocity of the cone,   
 and  a nearly constant conductivity comes from the phonon scattering 
 on the Dirac cone.
The presence of the off-diagonal component ($\sigma_{xy}$), which is associated with the deviation of the tilting axis of the Dirac cone from 
the $k_x$ or $k_y$ axis, results in the principal axis with clockwise or anticlockwise rotation depending on the sign of $\sigma_{xy}$.

Finally, we compare our result with that of the experiment.
The temperature dependence of resistance (corresponding to the inverse of the conductivity) under hydrostatic pressure  shows a nearly constant behavior at high temperatures and a minimum at low temperatures,  
 whereas the minimum is invisible under  uniaxial pressure.\cite{Kajita_JPSJ2014} 
Our results show the broad maximum for $\sigma_y$ and 
 monotonic variation for $\sigma_x$.
 Although we obtain the qualitative coincidence,  
   the details of the correspondence between them,  e.g., 
  the direction of measurement, are needed   for the quantitative comparison.
  It also remains a future problem 
   to clarify   if  the conductivity 
  under pressure  suggests the presence of site potentials. 
The validity of our present calculation may be examined 
by the measurement of the deviation angle of the principal axis, i.e., 
 clockwise (anticlockwise).

\acknowledgements
We thank R. Kato for useful 
 discussions on the effect of the e--p interaction. 
 This work was supported by Grants-in-Aid for Scientific Research 
 from the Japan Society for the Promotion of Science (Grant No. JP18H01162), 
and by JST-Mirai Program (Grant No. JPMJMI19A1).


\appendix

\section{Matrix element of H}
The quantity $h_{i,j}$, which denotes 
 the matrix element of $H$
  and  is the same as 
 that of Ref.\citen{Katayama_EPJ}, is given as 
\begin{subequations}
\begin{eqnarray}
h_{12}(\bk) &=& a_3 + a_2 Y\;, \\
h_{13}(\bk) &=& b_3 + b_2 X\;, \\
h_{14}(\bk) &=& b_4 Y + b_1 XY\;,\\ 
h_{23}(\bk) &=& b_2 + b_3 X\;, \\
h_{24}(\bk) &=& b_1 + b_4 X\;, \\
h_{34}(\bk) &=& 2 a_1\;, \\
h_{11}(\bk) &=& t_{22}(\bk) = a_{1d} (Y + \bar{Y})\;, \\
h_{33}(\bk) &=&  a_{3d} (Y + \bar{Y})+ \tV_{\rm B}\;, \\
h_{44}(\bk) &= &a_{4d} (Y + \bar{Y})+ \tV_{\rm C}\;, 
\end{eqnarray}
\end{subequations}
and
$h_{ij}(\bk) = h_{ji}^*(\bk)$, 
 where   
$X=\exp[i kx] = \bar{X}^*$  and  $Y= \exp[i ky] = \bar{Y}^*$, 

The matrix elements of $H_1$ are 
$\tV_B$ for $\alpha=\beta=3$,  $\tV_C$ for $\alpha=\beta=4$,  
and zero otherwise.
The site potentials  $\tV_{\rm B}$ and  $\tV_{\rm C}$ in $H_1$   
 are given by the mean--field of  
 short-range  repulsive interactions,~\cite{Katayama_EPJ}   
\begin{subequations}
\begin{eqnarray}
\label{eq:VB}
\tV_{\rm B} &=& (n_B-n_A)U/2  \nonumber \\
     &+ &  2 V_a(n_C-n_A) + 2V_b(2n_A -n_B-n_C) \; ,\nonumber \\
\\
\label{eq:VC}
\tV_{\rm C} &=& (n_C-n_A)U/2   \nonumber \\
     &+ &  2 V_a(n_B-n_A) + 2V_b(2n_A -n_B-n_C)\; , \nonumber \\
 \end{eqnarray} 
\end{subequations} 
 where 
  $U$ is the on-site repulsive interaction and   $V_a$ ($V_b$) 
 denotes the nearest neighbor interaction along the $y$ ($x$) axis.   
$n_{\alpha}$ denotes a local density corresponding to 
 an electron number per unit cell 
 at the $\alpha$ site and is determined self-consistently. 
In the present case,  the site potentials 
given by Eqs.~(\ref{eq:VB}) and (\ref{eq:VC})
are  estimated as  $\tV_B =0.0511$ and   $\tV_C = 0.0032$, respectively, 
 for 
$U$= 0.4, $V_a$=0.17, and  $V_b$=0.05, where   
 $n_A=n_{A'}$ = 1.46,   $n_B$ = 1.37, and   $n_C$ = 1.71, 
  and $\mu$ = 0.172 at $T$=0 .

\section{Damping by phonon scattering}
The damping of electrons of the $\g$ band, which is defined by 
 $\Gamma_\g$,  is obtained from the electron Green function\cite{Abrikosov} 
 expressed as   
\begin{subequations}
\begin{eqnarray}
 G_\g(\bk, i \omega_n)^{-1} & = & 
 i \omega_n - E_{\g,\bk}+ \mu 
  + i \Gamma_{\g}  
  \; ,
 \label{eq:eq14a} \\
\Gamma_{\g} & = & \Gamma + \Gamma_{\rm ph}^{\g}
 \; , 
 \label{eq:damping} 
  \end{eqnarray} 
\end{subequations}
 where 
$\Gamma_{\rm ph}^{\g} = - {\rm Im} \Sigma_\g (\bk, E_{\g, \bk} - \mu)$ 
 with $ \Sigma_\g (\bk, E_{\g, \bk} - \mu)$
 being  a self-energy given by the e--p interaction. 
 The real part of the self-energy 
 can be neglected for  doping at low concentrations.~\cite{Suzumura_PRB_2018}
 The quantity $\Gamma$ comes from another  self-energy by 
the impurity scattering.
Note that $\Gamma_{\rm ph}^\g$ does not depend on $\Gamma$,
 and that the ratio $\Gamma_{\rm ph}^{\g}/\Gamma$ 
  is crucial to the determination of  
 the $T$ dependence of the conductivity.    
The quantity $\Sigma_\g (\bk, \omega) = \Sigma_\g (\bk, i \omega_n)$ 
 with $i\omega_n \rightarrow \omega +  0$ 
 is  estimated as 
\cite{Abrikosov} 
\begin{eqnarray}
 & & \Sigma_\g (\bk, i \omega_n)  =  T \sum_m \sum_{\bq}\; |\alpha_q|^2 
            \nonumber \\
 & &\times   \frac{1}{i \omega_{n+m} - \xi_{\g, \bk+\bq}} 
      \times \frac{2 \omega_{\bq}}{\omega_{m}^2 + \omega_{\bq}^2} \; , 
 \label{eq:self_energy}
  \end{eqnarray} 
which is a product of electron and phonon Green functions. 
$\omega_n=  (2n+1)\pi T$, $\omega_{m}=2\pi m T$ with $n$ and $m$ being integers. $\xi_{\g, \bk} = E_{\g, \bk} - \mu$. 
 Applying the previous result,\cite{Suzumura_PRB_2018}
 we obtain Eqs.~(\ref{eq:eq16a}) and (\ref{eq:eq16b}).


\end{document}